\RequirePackage[2020-02-02]{latexrelease}
\documentclass[%
nofootinbib,
nobibnotes,
 amsmath,amssymb,
 aps
prd,
]{revtex4-2}
\usepackage[dvipsnames]{xcolor}
\usepackage{graphicx}
\usepackage{dcolumn}
\usepackage{bm}
\usepackage{hyperref}
\usepackage{graphicx}
\usepackage{geometry}
\usepackage{tikz}
\usetikzlibrary{decorations.pathmorphing}

\usepackage{bm}
\DeclareMathOperator{\sgn}{sgn}

\newcommand{\ham}{\mathcal{H}}
\newcommand{\diff}{\mathcal{D}}
\newcommand{\shift}{{N^x}}
\newcommand{\lapse}{N}
\newcommand{\erad}{E^x{}}
\newcommand{\ephi}{E^\varphi{}}
\newcommand{\krad}{K_x}
\newcommand{\kang}{K_\varphi}

\newcommand{\arguments}{r}

\def\exterior{E}
\def\interior{I}

\def\uint{\bar u}
\def\vint{\bar v}
\def\rint{\bar r}
\def\rEF{\tilde\rho}
\def\trianglephi{\Lambda}

\def\horizon{\mathcal{Z}}
\def\xu{z}
\def\time{t}
\def\radi{x}
\def\text{\tilde t}
\def\rext{\tilde r}
\def\timeI{\overline T}
\def\xuI{\overline Y}

\def\signgeod{\varepsilon}

\def\timeI{\overline T}
\def\xuI{\overline Y}

\def\propertime{s}

\begin{document}

\title{Nonsingular spherically symmetric black-hole model with holonomy corrections}

\author{Asier Alonso-Bardaji}
  \email{asier.alonso@ehu.eus}
	\affiliation{Department of Physics and EHU Quantum Center, University of the Basque Country UPV/EHU,
	Barrio Sarriena s/n, Leioa, Spain}

\author{David Brizuela}
  \email{david.brizuela@ehu.eus}
  	\affiliation{Department of Physics and EHU Quantum Center, University of the Basque Country UPV/EHU,
	Barrio Sarriena s/n, Leioa, Spain}
	
\author{Ra\"ul Vera}
  \email{raul.vera@ehu.eus}
  	\affiliation{Department of Physics and EHU Quantum Center, University of the Basque Country UPV/EHU,
	Barrio Sarriena s/n, Leioa, Spain}

	\begin{abstract}
          We present a covariant model of a spherically symmetric black hole with corrections motivated by loop quantum gravity. The effective modifications, parametrized by a positive constant $\lambda$, are implemented through a canonical transformation and a linear combination of the constraints of general relativity,
          in such a way that the theory remains free of anomalies
          and general relativity is recovered for $\lambda=0$.
          In addition, the corresponding metric is constructed in a fully covariant way
          to ensure that gauge transformations
          on phase space correspond to coordinate changes.
          The solution for each  gauge choice provides a chart
            and corresponding line element of a spacetime solution whose geometry is unambiguously determined in terms of the parameter $\lambda$ and a constant of motion $m$.	For positive values of $m$, the solution is asymptotically flat and contains a globally hyperbolic black-hole/white-hole
          region with a 
          minimal spacelike hypersurface 
          that replaces the Schwarzschild singularity.
        The corresponding exterior regions are isometric
        and, in particular, allow the computation of the ADM mass.
        The procedure to obtain the global causal structure of the solution yields also its maximal analytic extension.
	\end{abstract}

\maketitle

\section{Introduction}
Singularities of general relativity (GR) are expected to disappear once a complete quantum description of gravity is achieved. For instance, a candidate such as loop quantum gravity predicts a quantized spacetime, which presumably mends those defects. However, as yet, we must content ourselves with effective models implementing the expected discrete quantum corrections and try to reconcile them with the continuous notion of diffeomorphism symmetry of GR. The accuracy shown by the effective polymerized homogeneous models when compared to the full quantum dynamics \cite{Ashtekar:2011ni,Agullo:2016tjh,Rovelli:2013zaa} 
suggests that this approach might also provide an accurate description of the main features of nonhomogeneous models quantized in the context of loop quantum gravity (see, for instance, the introductory reviews on loop quantum gravity \cite{Rovelli:2011eq, Ashtekar:2021kfp}).

Considering nonhomogeneous spacetimes, spherically symmetric vacuum is the most simple, though interesting, scenario (for a review on fundamental issues of black-hole horizons in loop quantum gravity we refer to \cite{Perez:2017cmj}).
Several studies have analyzed the polymerization of such models,
particularly focusing on the
interior of the black hole, whose geometry is described by a Kantowski-Sachs metric. Making use of the available effective techniques for homogeneous
spacetimes, it turns out that bouncing scale factors are a general feature of such models.
Nonetheless, this approach is only partially satisfactory. First,
it dismisses many factors arising from the lack of homogeneity and the possibility to provide a
description of the exterior region. Second, these bounces
do not necessarily imply a tunneling into a white hole. Therefore, control over the complete modified spacetime metric is necessary in order to be able to
perform a comprehensive geometrical analysis.

In particular, the description of the exterior static region involves several problems regarding the asymptotic flatness, the slicing independence and the matching at the horizon. There are recent proposals that address some of the mentioned problems \cite{BenAchour:2018khr,Gambini:2020nsf,Kelly:2020uwj,Bodendorfer:2019cyv,Ashtekar:2018lag,Ashtekar:2018cay,Ashtekar:2020ckv,Bodendorfer:2019cyv,Bodendorfer:2019nvy,Bouhmadi-Lopez:2020oia} and predict the formation of either an inner horizon or a spacelike transition surface towards a white hole.
However, every model so far violates the usual notion of covariance \cite{Bojowald:2020dkb,Bojowald:2020unm}, particularly when introducing matter fields \cite{Bojowald:2015zha,Alonso-Bardaji:2020rxb}. The use of self-dual variables has been suggested as a possible way out for these no-go results \cite{Bojowald:2019fkv} but we will work with real Ashtekar-Barbero variables and, more precisely, with their spherical reduction \cite{Bojowald:2004af,Bojowald:2005cb}. Some vacuum studies following the consistent constraint deformation approach predict the formation of an Euclidean region in the deep quantum regime \cite{Bojowald:2018xxu}, where the notion of causality would be lost. This result shows that \textit{a priori} geometry should not be assumed but rather derived from the modified phase space \cite{Bojowald:2016hgh,Bojowald:2018xxu,Bojowald:2020unm,Bojowald:2021isp,ALONSOBARDAJI2022137075}.

Following the anomaly-free polymerization presented in \cite{Alonso-Bardaji:2021tvy} both for vacuum and matter fields with local degrees of freedom, in this paper
we study the physical implications of such model in the vacuum case,
extending and completing the results announced in \cite{ALONSOBARDAJI2022137075}.
We show, in particular,
  that the polymerization, which can be achieved by performing a certain canonical
  transformation and linear combination of classical constraints,
  can be naturally associated to a geometry, 
  ensuring that gauge transformations on phase space
  correspond to coordinate transformations on the spacetime solution. This is,
  different gauge choices on phase space lead to different charts
  of the same spacetime.

As it will be detailed below, see Fig.~\ref{diagram:maxextension}, the resulting spacetime
  replaces the Schwarzschild singularity with a minimal spacelike hypersurface inside the horizon,
  placed between a trapped and antitrapped regions. This interior domain
  thus consists of a black-hole region that smoothly
  emerges into a white-hole region.
  The exterior asymptotically flat regions are all isometric, and
  the fall-off properties of the Ricci tensor allow the computation
  of different global and quasilocal versions of mass,
  in particular the ADM mass. As a result all exterior regions
  are of equal mass.

{The article is organized as follows. In Sec.~\ref{sec:polham}, we introduce the effective Hamiltonian. In Sec.~\ref{sec.lineelements}, we construct the metric associated to {that} modified Hamiltonian. Then, in Sec.~\ref{sec:differentdomains}, we compute the explicit form of the metric in different charts that correspond to different gauge choices on phase space. We obtain a static region (Sec. \ref{sec:static}), a homogeneous region (Sec. \ref{sec:homogeneous}) and three domains (Sec. \ref{sec:coverings}) that cover the black-hole horizon and the transition surface between the black-hole/white-hole regions. In Sec.~\ref{sec:global}, we take one of the domains, which covers all the rest, and use it to study the global structure of the spacetime solution, which eventually yields the maximal analytic extension (see Sec.~\ref{sec:maxextension}). In Sec.~\ref{sec:properties}, we point out some of the relevant geometrical and physical properties of the model. More precisely, we study the causal structure (Sec. \ref{sec:trapped}), the curvature (Sec. \ref{sec:curvature}), some particular concepts of mass (Sec. \ref{sec:mass}), measurable effects outside the horizon (Sec. \ref{sec:qnm}), and the Schwarzschild and Minkowski limits (Sec. \ref{sec:limits}). We end the paper with a brief summary and discussion of the results in Sec.~\ref{sec:conclusion}.}

\section{The polymerized Hamiltonian}\label{sec:polham}

In order to construct the canonical formulation of GR
we start with a time-oriented manifold $M$ foliated by the spacelike level surfaces
of a given time function $t$. The usual analysis shows then that the Hamiltonian of general
relativity turns out to be a linear combination of four constraints and thus vanishes on shell. 
These constraints are the so-called Hamiltonian and diffeomorphism constraints.

In spherical symmetry one can introduce yet another function $\radi$ (such that
  $d\radi$ vanishes nowhere)
  on $M$ to be constant on the orbits of the spherical symmetry group.
Outside the fixed points of the group, $x$
  defines a radial direction. 
By choosing these adapted coordinates (and the usual angular coordinates on the symmetry orbits),
the angular components of the diffeomorphism constraint are
trivially vanishing and, in terms of Ashtekar-Barbero variables, its radial component reads as follows,
\begin{align}
\diff=
     -({\widetilde{E}^x})'{\widetilde{K}_x} +{\widetilde{E}^\varphi} ({\widetilde{K}_\varphi})',     
     \end{align}
where the prime stands for the derivative with respect to $x$.
On the other hand, the Hamiltonian constraint takes the form
     \begin{align}
    & \widetilde{\mathcal{H}}
       = -\frac{{\widetilde{E}^\varphi}}{2\sqrt{{\widetilde{E}^x}}}\left(1+\widetilde{K}_\varphi^2\right)  -2\sqrt{{\widetilde{E}^x}}{\widetilde{K}_x}{\widetilde{K}_\varphi}
       +\frac{1}{2}\Bigg(\frac{\widetilde{E}^x{}'}{2\widetilde{E}^\varphi}\left(\sqrt{\widetilde{E}^x}\right)'
              +\sqrt{\widetilde{E}^x}\left(\frac{\widetilde{E}^x{}'}{\widetilde{E}^\varphi}\right)'\Bigg)
\end{align}
where $\widetilde{E}^x>0$ is chosen to define
the positive orientation of the triad.
The model is completely described in terms of four dynamical variables: two independent components
of a densitized triad, $\widetilde{E}^x$ and $\widetilde{E}^\varphi$, and their respective conjugate momenta,
$\widetilde{K}_x$ and  $\widetilde{K}_\varphi$. Their Poisson brackets are given by the canonical forms,
$$\{\widetilde{K}_x(x_a),\widetilde{E}^x(x_b)\}=\delta(x_a-x_b), \qquad \{\widetilde{K}_\varphi(x_a),\widetilde{E}^\varphi(x_b)\}=\delta(x_a-x_b).$$
Concerning the algebraic structure, the above constraints form the algebra
\begin{subequations}\label{hdaclass}
  \begin{align}
    \big\{D[f_1],D[f_2]\big\}&=D\big[f_1f_2'-f_1'f_2\big],\\
    \big\{D[f_1],\widetilde{H}[f_2]\big\}&=\widetilde{H}\big[f_1f_2'\big],\\
    \label{hhbrackets}
    \big\{\widetilde{H}[f_1],\widetilde{H}[f_2]\big\}&=D\left[
    {\widetilde{E}^x(\widetilde{E}^\varphi)^{-2}}(f_1f_2'-f_1'f_2)\right],
  \end{align}
\end{subequations}
where we have defined the smeared form of the constraints
$D[f]:=\int dx f\diff$ and $\widetilde{H}[f]:=\int dx f\widetilde{\ham}$.
This explicitly shows that they are first-class constraints and thus
generators of gauge transformations. More precisely, the diffeomorphism
constraint generates deformations within each spacelike leaf of the foliation,
whereas the Hamiltonian constraint generates deformations of the hypersurfaces (as a set). 
The bracket \eqref{hhbrackets}
is of particular relevance.
On the one hand,
as it will be explicitly detailed below, the structure function on the right-hand
side, i.e., $\widetilde{E}^x/(\widetilde{E}^\varphi)^2$, encodes the information about the geometry on each hypersurface and the spacetime signature.
On the other hand, it ensures that the set of three-dimensional spacelike hypersurfaces
can be embedded in the spacetime manifold providing a foliation \cite{Teitelboim73}.
In summary, the above hypersurface deformation algebra reflects the covariance of GR in this canonical setting,
in which the Hamiltonian is given by
the combination $\widetilde{H}[N]+D[\shift]$,
with the Lagrange multipliers $\lapse$ and $\shift$
being the lapse and shift of the 3+1 decomposition
respectively.
To be precise,
let us recall that given the functions $\time$ and $\radi$ on $M$,
endowed with a metric $g$, the lapse and shift are obtained as follows. The normal direction of the slices is
given by the unit timelike vector $n=-N\nabla t$ (where $\nabla t$ is the vector metrically associated to the form $dt$), with $N:=1/\sqrt{-g(\nabla t,\nabla t)}>0$ the lapse function. We also take $n$ as the representative
of the future-pointing direction.
The shift is then defined as $\shift:=-n(x)/n(t)$ and, in terms of the natural basis of vector fields from $\time$ and $\radi$,
  $\partial_t$ and $\partial_x$, we explicitly have
  $n=\lapse^{-1}\left(\partial_\time-\shift\partial_\radi\right)$.

The quantization prescribed by loop quantum gravity makes use of holonomies of the connections
and fluxes of the triads as elementary variables. Therefore, in order to construct
effective models that would describe some of the expected features of such a quantization,
one usually follows a polymerization procedure. In the context of the present symmetry-reduced model,
this amounts to the replacement of the
component $\widetilde{K}_\varphi$ by a certain periodic function $f(\widetilde{K}_\varphi)$
in the classical Hamiltonian.
Nonetheless, this deformation function $f$ can not be completely arbitrary if one does not wish to introduce anomalies, this is, to spoil the first-class nature of the constraints.
In the vacuum case it is well known how to choose this
deformation function in order to produce an anomaly-free
deformed theory \cite{Bojowald:2015zha,Bojowald:2018xxu}. 
But for dynamical spacetimes coupled to matter with local degrees of freedom, such a simple procedure does
not provide a closed algebra \cite{Bojowald:2015zha,Alonso-Bardaji:2020rxb}, and further modifications are needed.

In particular, in \cite{Alonso-Bardaji:2021tvy} a family of anomaly-free deformed Hamiltonian constraints was obtained
by performing a systematic study under the only assumption that those constraints should be quadratic
in radial derivatives of the basic variables. Each of the Hamiltonian constraints in that family 
forms an algebra with the classical diffeomorphism constraint
under the presence of a scalar matter field, and thus also
trivially in vacuum. As explained in \cite{Alonso-Bardaji:2021tvy}, the anomaly-free requirement still leaves much freedom
left in the modified Hamiltonian constraint in the form of unspecified free functions.
Nevertheless, by following
a ``least deformation'' principle, a specific subfamily of Hamiltonian constraints was
obtained and characterized by a free function
$f(\widetilde{K}_\varphi)$.
That ``least deformation'' principle consists on requesting that the modified Hamiltonian remains in
form as close as possible to its GR counterpart while retaining some freedom that could
be interpreted as a holonomy correction. 
It turns out that this subfamily of Hamiltonian constraints
can also be obtained by performing the canonical transformation
of the classical variables presented in \cite{cantransf}, followed by a regularization procedure considered in \cite{Alonso-Bardaji:2021tvy}.
For the sake of brevity, we introduce next the construction of the deformed constraint in this latter way.

First the following canonical transformation is performed,
\begin{align}\label{cantransf}
     \widetilde{E}^x\rightarrow \erad\,,\quad 
     \widetilde{K}_x\rightarrow \krad\,,\quad
     \widetilde{E}^\varphi\rightarrow \frac{\ephi}{\cos(\lambda \kang)}\,,\quad \widetilde{K}_\varphi\rightarrow \frac{\sin(\lambda \kang)}{\lambda}\,,
\end{align}
for some real $\lambda\neq 0$,
which leaves the $(E^x,K_x)$ pair invariant. This transformation introduces a trigonometric function of $K_\varphi$, as expected for holonomy corrections, and amounts
to a specific choice of the free function mentioned above.
Note that, as opposed to the usual polymerization procedure, not only the variable $K_\varphi$
but also its conjugate $E^\varphi$ is changed by the above transformation in order to ensure
that it is canonical and thus the same constraint algebra is guaranteed.
The parameter $\lambda$, which can be taken to be positive for convenience
  without loss of generality,
is a dimensionless constant that, in this interpretation, is related to the fiducial length of the holonomies.
In particular, in the $\lambda\rightarrow 0$ limit the transformation is the identity and this will
be the limit where one recovers GR.
In addition, this canonical transformation is bijective as long as $\cos(\lambda {\kang})$ does not vanish. Hence, this transformation introduces the boundaries
$\cos(\lambda K_\varphi)=0$ on the classical phase space, which separate regions
where the dynamical trajectories can be mapped one to one to those given by GR
\cite{Bojowald:2021isp}.
Concerning the form of the constraints in terms of these new variables, the functional form of the
diffeomorphism constraint is unaltered, i.e.,
\begin{equation}\label{eq:D}
  {\cal D}= -{\erad}'{\krad} +{\ephi} \kang',
\end{equation}
whereas the Hamiltonian constraint does change and in particular gets some inverse terms
of $\cos(\lambda {\kang})$. Therefore, in order to regularize the poles $\cos(\lambda\kang)=0$, and include those surfaces
in our analysis, we consider the regularized constraint $\mathcal{C}:=\cos(\lambda {\kang})\widetilde{\cal H}$
along with its smeared form $C[f]:=\int f\mathcal{C}{dx}$. Even if vanishing on shell,
the Poisson bracket of this constraint with itself,
$$\big\{C[f_1],C[f_2]\big\}=D\left[\frac{\erad}{\ephi^2}\cos^4(\lambda\kang)(f_1f_2'-f_1'f_2)\right] -C\left[\frac{\lambda\sqrt{\erad}\erad'}{4\ephi^2}\sin(2\lambda\kang)(f_1f_2'-f_1'f_2)\right],$$
produces an additional term as compared
to \eqref{hhbrackets}. However, it is possible to perform a linear combination
with the diffeomorphism constraint so that the algebra takes its canonical form (see \cite{Alonso-Bardaji:2021tvy} for more details). More precisely, we define the deformed
Hamiltonian constraint as
\begin{align}
  {\cal H}:=&\left(\widetilde{\ham}
              +\lambda\, \sin(\lambda K_\varphi) \frac{\sqrt{\erad}\erad{}'}{2(\ephi)^2}{\cal D}\right)
              \frac{\cos(\lambda \kang)}{\sqrt{1+\lambda^2}}
              \nonumber\\
  =&  -\frac{{\ephi}}{2\sqrt{{\erad}}\sqrt{1\!+\!\lambda^2}}\left(1+\frac{\sin^2(\lambda {\kang)}}{{{\lambda^2}}}\right) 
     -\sqrt{{\erad}}{\krad}\frac{\sin(2\lambda \kang)}{\lambda\sqrt{1\!+\!\lambda^2}}\left(1+\left(\frac{\lambda \erad'}{2\ephi}\right)^{\!2}\right)
     \nonumber\\
            & +\frac{\cos^2(\lambda \kang)}{2\sqrt{1+\lambda^2}}
              \bigg(\frac{\erad'}{2\ephi}\left(\sqrt{\erad}\right)'
              +\sqrt{\erad}\left(\frac{\erad'}{\ephi}\right)'\bigg),
              \label{H_normal}
\end{align}
along with its smeared form $H[f]\!:=\!\int\!\! f {\cal H} dx$. Note that this linear combination with phase-space dependent coefficients may produce modifications of the dynamics of the theory (see, e.g., \cite{Bojowald:2022rjp}). Nonetheless, an important property of this construction is that in the limit $\lambda\to0$ GR is recovered since both the canonical transformation \eqref{cantransf} and the linear combination above are identities.

The constraint
algebra takes the canonical form
\begin{eqnarray}
\{D[f_1], D[f_2] \} &=& D[f_1f_2'-f_1' f_2],\nonumber\\
  \{D[f_1], {H}[f_2] \} &=& {H}[f_1f'_2],\label{algebra}\\
   \{{H}[f_1], {H}[f_2] \} &=& D[F(f_1f_2'-f_1' f_2)],\nonumber
\end{eqnarray}
with the structure function
\begin{align}\label{betak}
F:=
    \frac{\cos^2(\lambda {\kang})}{1\!+\!\lambda^2}\left(1+\Big(\frac{\lambda {\erad}'}{2{{\ephi}}}\Big)^{\!2}\right)\frac{\erad}{(E^{\varphi})^2}.
\end{align}
This function is non-negative. As its classical counterpart, it vanishes at $E^x=0$,
where the singularity is located in Schwarzschild,
but also at the surfaces defined by $\cos(\lambda K_\varphi)=0$.
In order to obtain the geometry associated to this modified canonical system,
the transformation properties
of this structure function will be of particular relevance.
But, before detailing those properties,
let us introduce the following object,
\begin{align}\label{eq.masspol}
   m:= \frac{\sqrt{{{\erad}}}}{2}\left(1+\frac{\sin^2(\lambda {\kang})}{\lambda^2}-\left(\!\frac{\erad'}{2{\ephi}}\!\right)^{\!\!2}\!\cos^2(\lambda {\kang})\right) ,
\end{align}
which commutes onshell with the modified Hamiltonian $H[N]+D[N^x]$ and it is thus
a constant of motion.
This quantity can be obtained as a direct polymerization of the Hawking mass \cite{Alonso-Bardaji:2021tvy},
and can be interpreted as the mass of the model (see Sec.~\ref{sec:mass} below). In addition to its physical relevance,
the importance of this observable is that
it explicitly shows that the zeros of the structure function $F$ given by the vanishing of $\cos(\lambda K_\varphi)$ are indeed covariantly defined. Note that,
since $K_\varphi$ is not a scalar quantity,
i.e., it does not define a function on $M$,
in general $\cos(\lambda K_\varphi)=0$ is a gauge-dependent
  condition and therefore does not covariantly define surfaces on $M$.
However, if one introduces this condition in the above definition
of mass \eqref{eq.masspol},
it is straightforward to obtain that $\cos(\lambda K_\varphi)=0$ if and only if $\sqrt{E^x}=2m\lambda^2/(1+\lambda^2)$,
which is a gauge-independent
statement because $E^x$ is a scalar.
In addition, looking at this last relation, it is natural to
introduce the new length scale
\[r_0:= 2m\frac{\lambda^2}{1+\lambda^{2}},\] which will define 
a minimum area $r^2_0$ of the model (see below).
Furthermore, in terms of $r_0$
the structure function $F$ takes the simpler expression
\begin{align}\label{F_explicit}
     F=\left(1-\frac{r_0}{\sqrt{{\erad}}}\right)\frac{\erad}{(\ephi)^2}.
\end{align}

In summary, we have introduced our modified Hamiltonian constraint by performing
the canonical transformation \eqref{cantransf} and the regularization \eqref{H_normal}.
The canonical transformation introduces trigonometric functions of
$\kang$ in the constraints, as expected from
loop quantum gravity, but also imposes artificial boundaries $\cos(\lambda\kang)=0$
in the classical phase space, which can not be described in
terms of the new variables. In this respect, the regularization achieves
two important goals. First, it removes the pole of the Hamiltonian
constraint on these surfaces, and thus allows to describe dynamical
trajectories crossing these boundaries. Second, the
structure function $F$ of the corresponding modified algebra has
just the right transformation properties in order to provide the system
with a geometrical interpretation, as we will show explicitly in the next section.

To end this section, we write explicitly the equations of motion of the system, which are given
by the Poisson brackets of the different variables with the Hamiltonian, 
\begin{subequations}\label{eom}
\begin{align}
\label{eomq1}
  \dot{\erad}=&\{\erad,D[\shift]+H[\lapse]\}= \shift \erad'+\lapse\sqrt{{\erad}}\frac{\sin(2\lambda \kang)}{\lambda\sqrt{1\!+\!\lambda^2}}\left(1+\left(\frac{\lambda \erad'}{2\ephi}\right)^{\!2}\right),
\\\label{eomq2}
    \dot{\ephi}=&\{\ephi,D[\shift]+H[\lapse]\}= \left(\shift \ephi\right)' +2\lapse\sqrt{\erad}\krad\frac{\cos(2\lambda \kang)}{\sqrt{1\!+\!\lambda^2}}\left(1+\left(\frac{\lambda \erad'}{2\ephi}\right)^{\!2}\right)\nonumber\\
              &+\lapse\frac{\sin(2\lambda \kang)}{\lambda\sqrt{1\!+\!\lambda^2}}\left(\frac{\ephi}{2\sqrt{\erad}} +
                \frac{\lambda^2}{2}\bigg(\frac{\erad'}{2\ephi}\left(\sqrt{\erad}\right)'
              +\sqrt{\erad}\left(\frac{\erad'}{\ephi}\right)'\bigg)
                \right),
\\\label{eomp1}
  \dot{\krad}=&\{\krad,D[\shift]+H[\lapse]\}= \left(\shift \krad\right)' +\lapse''\frac{\sqrt{\erad}  \cos ^2(\lambda \kang)}{2 \sqrt{1\!+\!\lambda^2} \ephi}\nonumber \\
              &+ \frac{\lapse'\sqrt{\erad}}{2\sqrt{1\!+\!\lambda^2}\ephi^2} \Bigg(\lambda\sin (2 \lambda \kang)\left(\erad'\krad-2\ephi\kang'\right) +\cos ^2(\lambda \kang)\left(\frac{\ephi\erad'}{2 {\erad} }-\ephi'\right)\!\!\Bigg)\nonumber\\
              &+\frac{\lapse}{\sqrt{1\!+\!\lambda^2}} \Bigg(\frac{\ephi (\sin ^2(\lambda \kang)+\lambda^2)}{4 \lambda^2  \erad^{3/2}} +\frac{\cos^2(\lambda\kang)}{4\sqrt{\erad}\ephi}\left(\erad''-\frac{(\erad')^2}{4\erad}-\frac{\erad'\ephi'}{\ephi}\right)\nonumber\\
              &-\frac{\krad\sin(2\lambda\kang)}{2\lambda\sqrt{\erad}}\left(1+\left(\frac{\lambda\erad'}{2\ephi}\right)^2\right)-\left[\sin(2\lambda\kang)\frac{\lambda\sqrt{\erad}}{2\ephi^2}\diff\right]'\Bigg),
\\\label{eomp2}
    \dot{\kang}=&\{\kang,D[\shift]+H[\lapse]\}= \shift \kang' +\lapse' \frac{\sqrt{\erad}\erad'}{2\ephi^2}\frac{\cos^2(\lambda \kang)}{\sqrt{1\!+\!\lambda^2}} -\lapse \frac{\sin^2(\lambda \kang)+\lambda^2}{2\lambda^2\sqrt{{\erad}}\sqrt{1\!+\!\lambda^2}}\nonumber\\
    &+\lapse\frac{(\erad')^2}{8\sqrt{\erad}\ephi^2}\frac{\cos^2(\lambda \kang)}{\sqrt{1\!+\!\lambda^2}} -\lapse\frac{\sin(2\lambda \kang)}{\sqrt{1\!+\!\lambda^2}}\frac{\lambda \sqrt{\erad}\erad'}{2\ephi^3}\diff,
\end{align}
\end{subequations}
in combination with the constraint equations $\diff=0$ and $\ham=0$.

\section{The construction of the metric}
\label{sec.lineelements}

As explained at the beginning of the previous section,
let us recall that we are given a time-oriented spherically symmetric manifold  $M$
  foliated by the function $t$ with spherically symmetric
  spacelike level hypersurfaces, and that we use
  $\radi$ (with $d\radi\neq 0 $) to denote some function which is constant on the orbits of the
  spherical symmetry group. 
  We want to endow $M$ with a metric tensor $g$. 
  We can use the pair $(\time,\radi)$ to produce a chart
  (that we denote in the following by $\{t,x\}$\footnote{The symbol $\{\cdot,\cdot\}$, which from now on will denote particular coordinate charts (ignoring the angular part),
  is not to be mistaken with the Poisson brackets in the previous section.})
on some domain in $M$
such that $g$ is given by the line element
\begin{equation}\label{sphericalmetric}
 ds^2=-L(t,x)^2 dt^2+q_{xx}(t,x)(dx+S(t,x) dt)^2+q_{\varphi\varphi}(t,x)d\Omega^2,
\end{equation}
where $L$ and $S$ are, respectively, the lapse and the radial
component of the shift vector of the usual 3+1 decomposition,
and $d\Omega^2$ denotes the metric of the unit sphere.

The goal is to reconstruct this metric
from the objects defined on the modified phase space. That is,
one needs to express the metric quantities $L$, $S$, $q_{xx}$,
and $q_{\varphi\varphi}$ in terms of $N$, $N^x$, $E^x$, $E^\varphi$,
$K_x$, and $K_\varphi$. For such a purpose, we will impose two natural conditions:
\begin{itemize}
\item[\textit{i/}]
The functions $t$ and $x$ define the Lagrange multipliers of the new
Hamiltonian in the same way as they do in GR.

\item[\textit{ii/}] Gauge transformations on the phase space
correspond to coordinate changes (over the corresponding domains) in the spacetime manifold.
\end{itemize}
Following the derivation of the lapse and shift from $t$ and $x$
in the GR case (see above)
the first condition is equivalent to requiring
that the lapse and the shift obtained from \eqref{sphericalmetric}
correspond to the Lagrange multipliers of the Hamiltonian,
that is $L(t,x)=N(t,x)$ and $S(t,x)=N^x(t,x)$ as functions on the
image of the chart $\{t,x\}$ of $M$. 

The second condition is more difficult to implement \cite{Teitelboim73,Pons:1996av}, so
let us analyze in detail this requirement by following
the developments in \cite{Bojowald:2018xxu}.

On the one hand, if we perform an infinitesimal
change of coordinates $(t+\xi^t,x+\xi^x)$, each component of the
line element \eqref{sphericalmetric} transforms as follows,
\begin{eqnarray}
\delta N &=& \dot{N} \xi^t+N' \xi^x+N (\dot{\xi}^t-N^x \xi^{t\prime}), \label{coordN}\\
  \delta N^{x} &=& \dot{N}^x \xi^t+N^{x\prime} \xi^x+N^x
    (\dot{\xi}^t-\xi^{x\prime})-\left[\frac{N^2}{q_{xx}} +(N^x)^2\right] \xi^{t\prime}+\dot{\xi}^x, \label{coordNx}\\
 \delta q_{xx} &=& \dot{q}_{xx} \xi^t+q_{xx}' \xi^x+2q_{xx} (N^x  \xi^t+\xi^{x\prime}),\label{coordqxx}\\
 \delta q_{\varphi\varphi} &=&\dot{q}_{\varphi\varphi} \xi^t+q_{\varphi\varphi}' \xi^x,
\label{coordqphiphi}
 \end{eqnarray}
which simply correspond to the components of the Lie derivative of
the metric along the vector $\xi=\xi^t\partial_t+\xi^x\partial_x$.

On the other hand, since they are first-class constraints,
${\cal H}$ and ${\cal D}$ are gauge generators. In particular,
the gauge transformation of any phase-space function 
$G=G(E^x,E^\varphi,K_x,K_\varphi)$ is given by the Poisson bracket
$\delta_\epsilon G=\{G,H[\epsilon^0]+D[\epsilon^x]\}$,
where $\epsilon^0$ and $\epsilon^x$ are the gauge parameters.
Thus, concerning the variables $(E^x, E^\varphi, K_x, K_\varphi)$, their gauge transformation
can be directly computed by these Poisson brackets. In fact, in the same
way as in GR, the time evolution is just a gauge transformation parametrized by the lapse $N$
and shift $N^x$. Therefore, one can immediately read the gauge
transformation of each of these phase-space variables from their
evolution equations \eqref{eom} simply by performing the replacements
$N\rightarrow\epsilon^0$ and $N^x\rightarrow\epsilon^x$.
However, as they are not phase-space variables,
obtaining the expression for the gauge transformation of
the Lagrange multipliers $N$ and $N^x$ is not so straightforward.
One can enlarge the phase space, as in \cite{Pons:1996av}, or analyze
the covariance of the equations generated by $H[N]+D[N^x]$ by
an explicit computation (see \cite{bojowald_2010}).
In either way,
one obtains that the gauge transformation of these multipliers
is given by
\begin{eqnarray}\label{gaugetrans1}
 \delta_\epsilon N&=&\dot{\epsilon}^0+\epsilon^x\partial_x N- N^x\partial_x\epsilon^0,\\\label{gaugetrans2}
  \delta_\epsilon N^x&=& \dot{\epsilon}^x+\epsilon^x N^{x\prime}-N^x \epsilon^{x\prime}
  -F(N\epsilon^{0\prime}-\epsilon^0 N').
\end{eqnarray}

Now, the gauge transformation parametrized by $\epsilon^0$ and $\epsilon^x$
should coincide with the coordinate transformation (the Lie
dragging) along the vector $\xi$ if those parameters correspond to the
normal decomposition of this vector, that is, 
$\xi=\epsilon^0 n +\epsilon^x \partial_x$
with $n=(\partial_t-N^x\partial_x)/N$
being the future-pointing unit normal to the hypersurfaces of constant $t$.
This condition implies the relations between components $\epsilon^0=N\xi^t$
and $\epsilon^x=\xi^x+\xi^t N^x$. Writing
\eqref{gaugetrans1}--\eqref{gaugetrans2}
in terms of the components of the generator of spacetime deformations,
\begin{eqnarray}\label{gaugeN}
 \delta_\epsilon N&=&\dot{N} \xi^t+N'\xi^x+N(\dot{\xi^t}- N^x \xi^{t\prime}),
 \\\label{gaugeNx}
  \delta_\epsilon N^x&=&\dot{N}^x\xi^t+ N^{x\prime}\xi^x
  + N^x(\dot{\xi}^t-\xi^{x\prime})
  -\left[F N^2+(N^x)^2\right] \xi^{t\prime}+ \dot{\xi^x},
\end{eqnarray}
it is then clear that these gauge transformations are identical to the above
coordinate transformations \eqref{coordN}--\eqref{coordNx}
if one defines $q_{xx}=1/F$. In order to check whether this is consistent, one also needs to analyze the transformation properties of $1/F$ with $F$
defined in \eqref{F_explicit}. The bracket $\{1/F, H[\epsilon^0]+D[\epsilon^x]\}$ gives the gauge transformation
\begin{align}
 \delta_\epsilon(1/F)&={ \frac{\epsilon^0}{N}(1/F){\bm{\dot{}}}\, + \left(\epsilon^x-\frac{N^x}{N}\epsilon^0\right) (1/F)'  +2\left(\epsilon^{x\prime}-\frac{\epsilon^0}{N} N^x{}'\right)(1/F)}\nonumber\\
 &= (1/F){\bm{\dot{}}}\,\xi^t + (1/F)'\xi^x +2{(1/F)}(N^x\xi^{t\prime} +\xi^{x\prime}),
\end{align}
after using the equations of motion \eqref{eom} and the relation between the components $\epsilon^0=N\xi^t$ and $\epsilon^x=\xi^x+\xi^t N^x$ given above.
Comparing this last expression with the transformation of $q_{xx}$ \eqref{coordqxx}, it is then clear that
the choice $q_{xx}=1/F$ is consistent and will lead to a covariant line element.

Finally, we are just left with the angular component of the metric $q_{\varphi\varphi}$. This object, as \eqref{coordqphiphi} shows, transforms
as a scalar, so any scalar phase-space function could be chosen to define it. Nonetheless,
in order not to introduce any strictly necessary deformation, 
and taking into account that the variable $\erad$ has remained unaltered
in the process of construction of the deformed Hamiltonian constraint,
we will consider $q_{\varphi\varphi}$ to have its classical
form $q_{\varphi\varphi}=\erad$. This can be equally argued by demanding
that in our new setting the phase-space function $\erad$ determines,
as in GR,
  the area of the orbits of the spherical symmetry group by $4\pi \erad$,
  that is, we keep $\sqrt{\erad}$ being the intrinsically defined area-radius function.
To sum up, this requirement on $\erad$ plus conditions $i/$ and $ii/$ above
leads to the construction of the metric tensor
\begin{equation}\label{metric}
 ds^2=-\lapse(t,x)^2 dt^2+{\frac{1}{F(t,x)}}(dx+\shift(t,x) dt)^2+\erad(t,x) d\Omega^2,
\end{equation}
which, after introducing \eqref{F_explicit}, reads explicitly
\begin{equation}\label{metric_orig}
    {ds}^2 = -\lapse(t,x)^2{d\time}^2 +\left({1-\frac{r_0}{\sqrt{\erad(t,x)}}}\right)^{\!-1}\frac{\ephi(t,x)^2}{\erad(t,x)}{\big({dx}+\shift(t,x){dt}\big)^2}
    +\erad(t,x){d\Omega}^2.
\end{equation}

Alternatively, in Ref.~\cite{Bojowald:2018xxu} there is another
proposal for a line element
in the context of a related polymerized model, relaxing the condition {\it i/} above.
The choice presented in that paper can be derived by considering a scalar conformal factor
multiplying the last line element.
That way one could also arrive to a covariant line element in the sense of condition {\it ii/},
even if, for instance, the unit normal ceases to be unit with the deformed metric.
Nonetheless, 
the explicit model eventually constructed in \cite{Bojowald:2018xxu}
turns out not to be covariant as the conformal factor considered there is not a scalar quantity,
as we show in App.~\ref{app.bby}.

\section{The solution in different domains}\label{sec:differentdomains}

In this section we start with
the chart 
  $\{\time,\radi\}$
(plus the angular coordinates) on some domain of $M$ in which the metric tensor $g$ is
given by the line element \eqref{metric_orig}.
The gauge freedom inherent to the
problem allows us to fix two relations (gauge choice) between the functions
on phase space.
For each gauge choice the solution to the system of equations
\eqref{eom} will yield a corresponding line element
of the form \eqref{metric_orig}.
Since our construction is consistent,
different gauge choices will simply lead to different
charts
(with different domains in $M$ in general) and corresponding
expressions (line elements) for the same metric,
thus providing a \emph{unique spacetime solution}.

In order to illustrate
how the gauge choices relate to the charts and the geometry,
thus checking the consistency of the construction,
and to find the global structure of the solution,
we will make five specific gauge choices, solve the system of equations
and produce the explicit form of the metric
in the respective charts on their domains. In each case we will relabel $t$ and $\radi$ to
distinguish the charts.
Before moving on to solve the system of equations, let us remark
that we will exclude degenerate solutions with identically vanishing $N$ or $\ephi$.

This section is divided in three subsections.
In Sec.~\ref{sec:static} we introduce a static gauge, which will provide
an exterior region. Section~\ref{sec:homogeneous} presents a homogeneous
gauge, which will describe (half of) the interior region. Finally, in Sec.~\ref{sec:coverings} we will consider three gauges that cover the previous cases. In particular, one of the three gauges
will produce a chart with a covering domain $\mathcal{U}$, which will be used in Sec. \ref{sec:global}
to analyze the global causal structure of the spacetime solution
and obtain its maximal analytic extension.

\subsection{Static region}
\label{sec:static}
We start by choosing the gauge condition
$\sin(\lambda\kang)=0$, so that $\sin(2\lambda\kang)=0$
  and $\cos(2\lambda\kang)=1$, which leads to the explicit form of the
equations of motion,
\begin{subequations}
\begin{align}
\dot{\erad}&= \shift \erad',\label{Nx_1}\\
  \dot{\ephi}&= \left(\shift \ephi\right)' +2\lapse\frac{\sqrt{\erad}\krad}{\sqrt{1\!+\!\lambda^2}}\left(1+\left(\frac{\lambda \erad'}{2\ephi}\right)^{\!2}\right),\\
    \dot{\krad}&= \left(\shift \krad\right)' +  \left(\frac{\lapse'\sqrt{\erad}}{2\sqrt{1+\lambda^2}\ephi}\right)'  +\frac{\lapse}{4\sqrt{1+\lambda^2}\sqrt{\erad} }  \left(\frac{\ephi}{\erad} + \frac{\erad''}{ \ephi} -\frac{(\erad')^2}{4  \erad\ephi}
     -\frac{\erad'\ephi'}{\ephi^2}
                 \right),\\
 0=\dot{\kang}&= \frac{\lapse'\sqrt{\erad}\erad'}{2\sqrt{1+\lambda^2}\ephi^2} -\frac{\lapse}{2\sqrt{1+\lambda^2}\sqrt{\erad}}\left(1-\frac{(\erad')^2}{4\ephi^2}\right) ,\\
    0=\diff &=- \krad\erad',\label{diff_constraint_1}\\
    0=\ham &=-\frac{{\ephi}}{2\sqrt{{\erad}}\sqrt{1+\lambda^2}}
             +\frac{1}{2\sqrt{1+\lambda^2}}
             \bigg(\frac{1}{2}\frac{\erad'}{\ephi}\left(\sqrt{\erad}\right)'
              +\sqrt{\erad}\left(\frac{\erad'}{\ephi}\right)'\bigg).
             \end{align}
\end{subequations}
If we further impose $\erad=\radi^2$, then
the vanishing of the diffeomorphism constraint \eqref{diff_constraint_1}
yields $x\krad=0$, while $\dot{\erad}=0$ demands $x\shift=0$ on \eqref{Nx_1}.
The remaining equations are
\begin{subequations}
\begin{align}
 0= \dot{\ephi}&,\label{ephizero}\\
    0=\dot{\krad}&=  \left(\frac{\lapse'\radi}{2\sqrt{1+\lambda^2}\ephi}\right)'  +\frac{\lapse}{4\sqrt{1+\lambda^2}\radi }  \left(\frac{\ephi}{\radi^2} + \frac{2}{ \ephi} -\frac{1}{\ephi}
     -\frac{2\radi\ephi'}{\ephi^2}
     \right),\label{kxdot}\\
    0=\dot{\kang}&= \frac{\lapse'\radi^2}{\sqrt{1+\lambda^2}\ephi^2} -\frac{\lapse}{2\sqrt{1+\lambda^2}\radi}\left(1-\frac{\radi^2}{\ephi^2}\right) ,\label{kphidot0}\\\label{hamstatic}
    0=\ham &=-\frac{{\ephi}}{2\radi\sqrt{1+\lambda^2}}
  +\frac{\radi}{2\sqrt{1+\lambda^2}\ephi}\bigg(3  -\frac{2\radi\ephi'}{\ephi} \bigg) ,
\end{align}
\end{subequations}
where $x$ has been chosen positive without loss of generality.
From \eqref{eq.masspol} we obtain,
\begin{align}
    \ephi =\varepsilon_0\radi\left(1-\frac{2m}{\radi}\right)^{-1/2},
\end{align}
with $\varepsilon_0^2=1$ and \eqref{ephizero} and \eqref{hamstatic} are thence automatically satisfied.
Therefore, the solution will only be valid for $\radi>2m$.
Now, we can integrate \eqref{kphidot0} for the lapse, so that
\begin{align}
    \lapse = c_1\sqrt{1-\frac{2m}{\radi}},
\end{align}
with a nonzero constant $c_1$.
One can check that the only remaining equation \eqref{kxdot}
is now automatically satisfied.

We now relabel  $(\time,\radi)$
as the pair of real functions $(\text,\rext )$ on $M$
that define a chart (in addition to the angular coordinate functions)
with ranges given by the domain of existence of the solutions.
Explicitly, the domain $D_S$ of the chart  $\Psi^{D_S}=\{\text,\rext\}$
on $M$
is only restricted by 
$\rext\in (2m,\infty)$.
The line element \eqref{metric_orig} in this chart
thus reads
\begin{align}\label{deformedschwarzsmetric}
    {ds}^2&=-\bigg(1-\frac{{2m}}{\rext}\bigg){d\tilde{t}}^2
    +\bigg(1-\frac{r_0}{\rext}\bigg)^{\!-1}\bigg(1-\frac{{2m}}{\rext}\bigg)^{\!-1}{d\rext}^2+\rext^2{d\Omega}^2,
\end{align}
where we have trivially absorbed $c_1$ into $\text$.
As it will be detailed below, this domain describes one exterior
(asymptotically flat) region.

\subsection{Homogeneous region}
\label{sec:homogeneous}
Now we 
start by demanding $\erad{}'=\ephi{}'=0$.
The vanishing of the diffeomorphism constraint $\mathcal{D}=0$,
cf. \eqref{eq:D}, implies $\kang'=0$. Then, the radial derivative
of the equation $\ham=0$, cf. \eqref{H_normal}, implies, in turn, that $\krad'=0$.
The radial derivatives of equations \eqref{eomq1} and \eqref{eomq2} therefore imply
$\lapse'=0$ and $\shift{}'=0$.
We can now partially use the gauge freedom left to impose
$\shift=0$.\footnote{Let us note that, from the spacetime perspective,
  this amounts to the fact
  that, because $\shift'=0$, there is a function $Y$ such that
  $d\radi+\shift(\time) d\time=dY$.}
The final form of the equations is then given by
\begin{subequations}
\begin{align}
\dot{\erad}&=\lapse\sqrt{{\erad}}\frac{\sin(2\lambda \kang)}{\lambda\sqrt{1+\lambda^2}},\label{lapse_hom}\\
\label{homephi}
    \dot{\ephi}&= \frac{\lapse}{\sqrt{1+\lambda^2}}\left(2\sqrt{\erad}\krad{\cos(2\lambda \kang)} +\frac{\ephi\sin(2\lambda \kang)}{2\lambda\sqrt{\erad}}\right),\\
    \dot{\krad}&=  \frac{\lapse}{2\sqrt{\erad}\sqrt{1+\lambda^2}} \left(\frac{\ephi }{2 \erad}\left(1+\frac{\sin^2(\lambda\kang)}{\lambda^2}\right) -\krad\frac{\sin (2 \lambda \kang)}{\lambda } \right),\\
    \dot{\kang}&=  -\frac{\lapse}{2\sqrt{\erad}\sqrt{1+\lambda^2}}\left(1+\frac{\sin^2(\lambda\kang)}{\lambda^2}\right),\\
    0=\ham &= -\frac{{\ephi}}{2\sqrt{{\erad}}\sqrt{1+\lambda^2}}\left(1+\frac{\sin^2(\lambda {\kang)}}{{{\lambda^2}}}\right) 
-\sqrt{{\erad}}{\krad}\frac{\sin(2\lambda \kang)}{\lambda\sqrt{1+\lambda^2}}.\label{H0_hom}
\end{align}
\end{subequations}
Interestingly, these are the same equations as those obtained with the usual
(anomaly-free) holonomy corrections \cite{BenAchour:2018khr}, up to the additional global
$\sqrt{1+\lambda^2}$ constant.
However, at this point it is important to note
that in this gauge a bouncing behavior is a general feature of some of the
individual variables, but it is not necessarily of the spacetime metric.
For instance, for $\lapse\sqrt{\erad}>0$ the points $2\lambda\kang=(2n+1)\pi$ with
integer $n$ are minima of $\erad$ since the right-hand side of \eqref{lapse_hom}
vanishes and one can show that $\ddot{\erad}>0$.
Conversely, $2\lambda\kang=2n \pi$ are maxima of $\erad$.
But such a statement is certainly gauge dependent and one needs to construct the metric
in order to extract the physical significance, if any, of those bounces.

In order to continue with such construction, we use the freedom left in the gauge choice
  at this point to set $\erad=\time^2$ and restrict the
  range of $\time$ to non-negative values $\time\geq0$.
  The constant of motion \eqref{eq.masspol} can now be used to obtain
  $\kang$ by
\begin{align}
  \frac{\sin(\lambda\kang)}{\lambda}  = \varepsilon_1\sqrt{\frac{2m}{t}-1},
  \label{ht1}
\end{align}
where $\varepsilon_1^2=1$, from where
\begin{align}
  \sin(2\lambda\kang)=2\varepsilon_2\lambda\sqrt{\frac{2m}{\time}-1}\sqrt{1-\lambda^2\left(\frac{2m}{\time}-1\right)}
  =2\varepsilon_2\frac{\sqrt{2m r_0}}{2m-r_0}\sqrt{\frac{2m}{\time}-1}\sqrt{1-\frac{r_0}{\time}},\label{ht2}
\end{align}
with $\varepsilon_2^2=1$.
In the last step (as we will do in the following)
we have used the definition of $r_0$. 
To ensure the existence of the solution,
relations \eqref{ht1} and \eqref{ht2}
impose in particular the restriction $r_0\leq t\leq 2m$.

With the above expressions, equation \eqref{lapse_hom} reads explicitly
\begin{align*}
 \varepsilon_2 \sqrt{\frac{2m}{t}-1}\sqrt{1-\frac{r_0}{t}}\lapse=1,
\end{align*}
that we solve for the lapse $\lapse$. Note that this solution
further restricts the range of $t$ to $r_0< t< 2m$.
From \eqref{homephi} we get
\begin{align*}
  \dot{\ephi}   =\frac{\ephi}{2}\frac{2m+r_0-2t}{(2m-t)(t-r_0)},
\end{align*}
which yields
\begin{align}
    \ephi  
  = c_2\sqrt{2m-t}\sqrt{t-r_0},
\end{align}
for some constant $c_2\neq 0$.
Finally, equation $\ham=0$, cf. \eqref{H0_hom}, allows us to obtain
\begin{align*}
   \krad =
   -\varepsilon_2\frac{\ephi}{4t^2} \frac{\sqrt{2m}\sqrt{2m-r_0}}{\sqrt{2m-t}\sqrt{t-r_0}}=-\varepsilon_2 c_2 \sqrt{2m}\sqrt{2m-r_0}\frac{1}{4t^2}.
\end{align*}

This time we relabel $(\time,\radi)$
  as a pair of real functions $(T,Y)$ on $M$
  so that the domain $D_h$ of the chart
  $\Psi^{D_h}=\{T,Y\}$ on $\mathbb{R}^2$ is restricted by the points where
$T\in (r_0,2m)$. The metric \eqref{metric_orig} in that chart, after absorbing $c_2$ in $Y$,
thus reads
\begin{align}\label{metschwarzshom}
    {ds}^2 &= -\bigg(1-\frac{r_0}{T}\bigg)^{-1}\!\bigg(\frac{{2m}}{T}-1\bigg)^{-1}{{dT}^2}
    +\bigg(\frac{{2m}}{T}-1\bigg){{dY}^2}+T^2{d\Omega}^2.
\end{align}
This domain corresponds to a Kantowski-Sachs (homogeneous)
type and will describe one half of an interior region, as it will be shown later.

\subsection{Covering domains}\label{sec:coverings}
The regions defined by the charts in the previous two subsections, $D_S$ and $D_h$,
cover different ranges of the area-radius function $\sqrt{E^x}$.
In particular,
the static region $D_S$ is defined for the range $(2m,\infty)$, whereas in the homogeneous
  region $D_h$ the function $\sqrt{E^x}$ takes values in the interval $(r_0,2m)$.
  As a result,  $D_h$ and $D_S$ do not intersect. Furthermore,
  $D_S$ does not cover the horizon that, as we will see,
  forms at the limit where $\rext \to 2m$, and $D_h$ does not cover the instant $T=r_0$.
The three gauge choices that we introduce in this subsection will
produce three charts on respective domains $D_H$,
$D_{EF}$ and $\mathcal{U}$.
The domain $D_H$ will cover two regions isometric to
  $D_h$. $D_{EF}$ will cover any region isometric to $D_S$ and
  $D_h$, while $\mathcal{U}$ will cover any region isometric to $D_{EF}$
and $D_H$.
In the next section
we will use the covering domain $\mathcal{U}$
to study the global causal structure of the spacetime solution
and obtain its maximal analytic extension.

\subsubsection{The whole homogeneous interior domain $D_H$}\label{sec:DH}

We start with the same requirement
  as in the homogeneous region, that is, $\erad'=\ephi'=0$.
We recall that this implies $\krad'=\kang'=\lapse'=0$
and that we can choose $\shift=0$.
The final choice we take now is to impose $\kang=t/\lambda$.
The solution of the system \eqref{lapse_hom}-\eqref{H0_hom}
for a suitable choice of constants of integration is then given by
\begin{align*}
  &\lapse=-\frac{\sqrt{1+\lambda^2}}{m\lambda}E^x,\quad 
    \erad=\left(\frac{2m\lambda^2}{\lambda^2+\sin^2t}\right)^2,\quad
    \ephi=-\frac{\lambda m \sin(2t)}{\sqrt{1+\lambda^2}(\lambda^2+\sin^2t)},\quad
    \krad=\frac{(\lambda^2+\sin^2 t)^2}{8m\lambda^4\sqrt{1+\lambda^2}},
\end{align*}
where $t$ is restricted by the roots of $\ephi$. Because of the periodicity of the solution, we can stick to the range $t\in (0,\pi)$ without loss of generality.
We relabel $(\time,\radi)$ as the pair of real functions $(\timeI,\xuI)$
on $M$
so that the domain $D_H$ of the chart $\Psi^{H}=\{\timeI,\xuI\}$ 
is the preimage of the stripe $\timeI\in(0,\pi)$ in $\mathbb{R}^2$.
After naming $\sqrt{\erad}=:\rint$, and using the definition of $r_0$,
the metric \eqref{metric_orig} in that chart thus reads 
\begin{equation}\label{g_I_full}
  ds^2=-\frac{2}{m r_0}\rint(\timeI)^4 d\timeI^2
  +\left(\frac{2m}{\rint(\timeI)}-1\right)
  d\xuI^2+\rint(\timeI)^2 d\Omega^2,
\end{equation}
with
\begin{equation}
  \rint(\timeI)
  =\frac{2mr_0}{r_0+(2m-r_0)\sin^2\timeI}\qquad \Longleftrightarrow\qquad
    \frac{2m}{\rint(\timeI)}-1=\left(\!\frac{2m}{r_0}-1\!\right)\sin^2\timeI.
  \label{r_T}
\end{equation}
This region contains the spacelike hypersurface $\rint=r_0$ located at
$\timeI=\pi/2$.
We will show later how this region covers two homogeneous regions $D_h$,
and that it describes the globally hyperbolic (in that sense, ``whole'')
homogeneous Kantowski-Sachs interior region of the solution.

\subsubsection{The covering domain $\mathcal{U}$}\label{sec:U}

For the next gauge choice we demand that $\dot{\erad}=0$ and $\dot{\ephi}=0$.
We start writing the diffeomorphism constraint $\mathcal{D}=0$, 
cf. \eqref{eq:D},
and 
the constant of motion \eqref{eq.masspol}
in explicit form as
\begin{subequations}
\begin{align}
  \erad'\krad&=\ephi\kang',\label{Kx_}\\
  \label{Kphi_}\cos^2(\lambda \kang)&=(1+\lambda^2)\left(1+\left(\frac{\lambda\erad'}{2\ephi}\right)^{2}\right)^{-1}\left(1-\frac{r_0}{\sqrt{\erad}}\right),
\end{align}
\end{subequations}
where we have used 
$\lambda^2=r_0/(2m-r_0)$ 
to remove $\lambda$ from the last factor.
We check first that if $\erad'=0$ then the vanishing of the Hamiltonian constraint \eqref{H_normal}, $\ham=0$,
implies that $\sin(2\lambda\kang)$ cannot vanish identically, and therefore
equation \eqref{eomq1} for $\dot\ephi=0$ yields $\lapse=0$.
We thus assume that $\erad'$ does not vanish identically in the following.
This, used in \eqref{Kphi_}, implies in particular that $\cos^2(\lambda\kang)$
  cannot vanish identically. Moreover, since we want to avoid
  the case $\sin(\lambda \kang)=0$, which has been treated already in Sec. \ref{sec:static},
we shall also assume in the following that $\sin(2\lambda\kang)$ does not vanish identically.

The solution for the lapse and shift is found
from the system of evolution equations \eqref{eomq1} and \eqref{eomq2}
with $\dot{\erad}=0$ and $\dot{\ephi}=0$ as follows.
We first isolate $\shift$ from \eqref{eomq1} and introduce it,
  together with \eqref{Kx_},
  into \eqref{eomq2}, from where we obtain
  \begin{equation*}
    \sin(2\lambda\kang)\erad\left(1+\left(\frac{\lambda \erad{}'}{2\ephi}\right)^2\right)
    \left(\lapse'\erad'\ephi+\lapse( \ephi'\erad'-\ephi\erad'')\right)=0.
  \end{equation*}
  Therefore,
\begin{equation}
  \lapse=\frac{c_4}{2}\,\frac{\erad'}{\ephi},
\end{equation}
for some non-zero constant $c_4$. Introducing this in \eqref{eomq1},
in combination with \eqref{Kphi_}, we obtain
\begin{equation}
    \shift= \varepsilon_3{c_4}\frac{\sqrt{\erad}}{\ephi} \sqrt{1-\frac{r_0}{\sqrt{\erad}}}\sqrt{\frac{2m}{\sqrt{\erad}}-1+\left(\frac{\erad'}{2\ephi}\right)^{\!2}},
\end{equation}
where $\varepsilon^2_3=1$ corresponds to minus
the sign of $\sin(2\lambda\kang)$.
It can be checked that all the remaining equations are now satisfied.
At this point let us denote, for compactness, $\erad$ and $\ephi$ as $\sqrt{\erad}=:r$ and $\ephi=:s$.
Note that this is just notation to describe the two free functions that are still to be fixed, and that we use $r$ to denote, as usual, the area-radius function.
In terms of these, the family of line elements 
takes the form,
\begin{align}\label{r-s-metric}
    {ds}^2 =&
    -\left(1-\frac{2m}{r(\radi)}\right) {d\time}^2
              +2\left(1-\frac{r_0}{r(\radi)}\right)^{-1/2}\frac{s(\radi)}{r(\radi)}
              \sqrt{\left(\frac{r(\radi) r'(\radi)}{s(\radi)}\right)^2+\frac{2m}{r(\radi)}-1}\,{dt}{d\radi}\nonumber\\
    &+ \left(1-\frac{r_0}{r(\radi)}\right)^{-1}\left(\frac{s(\radi)}{r(\radi)}\right)^{2} 
    {d\radi}^2 
   +r(\radi)^2 {d\Omega}^2 \:,
\end{align}
where we have set $\varepsilon_3 c_4=1$
with no loss of generality by a constant rescaling (and change of sign) of $t$.

Observe that the function $s(\radi)$ could be absorbed by performing
a change $s(\radi)d\radi \to dX$, but it is convenient to keep that
freedom to find convenient charts later.
The values of $\radi$ ought to be restricted, in particular,
by the form of the functions $r(\radi)$ and $s(\radi)$ so that the line element
\eqref{r-s-metric} and $K_\varphi$ [see \eqref{Kphi_}]
are well defined.
Next we will use the freedom left in $r(\radi)$ and $s(\radi)$ to produce two
different gauge choices, which will thence provide two charts.
This time the domains relative to these two charts
will present a nonempty intersection.

We want to find a chart that contains the points reaching the minimum $r=r_0>0$.
Since some of the divergences in the line element \eqref{r-s-metric}
come from $q_{xx}$ (the coefficient of $dx^2$),
let us try the choice
$s^2=r^2(1-r_0/r)$, so that $q_{xx}=1$.
Thus, the factor in front of $d\radi^2$ in \eqref{r-s-metric}, also
present with $d\time d\radi$, equals one,
and the only possible divergences  appear in the argument of the square root,
\begin{align}\label{squareroot}
    \left(1-\frac{r_0}{r(\radi)}\right)^{-1}r'(\radi)^2+\frac{2m}{r(\radi)}-1.
\end{align}
The option we take is to choose the following particular condition [after relabeling $(\time,\radi)$ to $(\tau,\xu)$],\footnote{We employ the usual definition of the sign function $\sgn$, so that
    as a function we use $\sgn(\xu)=0$ where $\xu=0$. Observe that it is not
    differentiable there. In a distributional
    sense, though, we use $\sgn(\xu)^2=1$, so that higher (even)
    derivatives of $r(\xu)$, as functions, attain their limiting values at $\xu=0$.
  }

\begin{equation}
  \label{rx}
  \frac{dr(\xu)}{d\xu}=\sgn(\xu) \sqrt{1-\frac{r_0}{r(\xu)}}
  \quad \mbox{ with }\quad r(0)=r_0,
\end{equation}
so that the argument of the square root \eqref{squareroot}
reduces to $2m/r$. Defined in this way,
$r(\xu)$ is an analytic function on $\mathbb{R}$
such that $r(-\xu)=r(\xu)$, it attains a minimum positive value $r_0=r(0)>0$ at $\xu=0$,
and it is given implicitly by
\begin{equation}
 \xu=\sqrt{r(\xu)}\sqrt{r(\xu)-r_0}+r_0\log\left(\sqrt{\frac{r(\xu)}{r_0}}+\sqrt{\frac{r(\xu)}{r_0}-1}\right),\quad\quad\mbox{for $z>0$}.
  \label{x_r}
\end{equation}
Observe that $r\to|\xu|$ as $\xu\to\pm\infty.$
With these choices we end up with a chart $\Psi^{\mathcal{U}}_{\tau,\xu}=\{\tau,\xu\}$
defined on some domain $\mathcal{U}$ in which the metric reads
\begin{align}\label{g:tau_x}
    \!\!{ds}^2&= 
    -\left(1-\frac{{2m}}{r(\xu)}\right) {d\tau}^2 
    +2\sqrt{\frac{{2m}}{r(\xu)}\,}\,{d\tau}{d\xu}  + {{d\xu}^2} +r(\xu)^2\, {d\Omega}^2\:.
\end{align}
The ranges of coordinates are given by $(\tau,\xu) \in \mathbb{R}^2$,
that is, the image of the domain $\mathcal{U}$ through the chart
$\Psi^{\mathcal{U}}_{\tau\xu}$ is the whole plane,
and the function $r$ as a function on $\mathcal{U}$,
that is $r:\mathcal{U}\to\mathbb{R}$,
is bounded from below by $r_0>0$.

\subsubsection{Eddington-Finkelstein-like domain}
\label{sec:EF}
Another, simpler, choice to impose on the free functions of the line element \eqref{r-s-metric}
is
to take $r(\radi)=\radi$ and $s(x)=x$.
After a convenient relabeling $(\time,\radi)$ as $(\tilde\tau,\rEF)$,
we thus obtain 
\begin{align}\label{metricglp}
    {ds}^2= 
    -\bigg(1-\frac{{2m}}{\rEF}\bigg) {d\tilde\tau}^2 +2\sqrt{\frac{2m}{\rEF-r_0}}\,{d\tilde\tau}{d\rEF}+ \bigg(1-\frac{{r_0}}{\rEF}\bigg)^{\!-1}{d\rEF}^2 
   +\rEF^2 {d\Omega}^2 \:.
\end{align} 
The range of the chart $\Psi^{EF}=\{\tilde\tau,\rEF\}$, defined on some domain
$D_{EF}\subset M$, is the half plane defined by $\rEF>r_0$.
Therefore this chart fails to describe the hypersurface $r=r_0$.
It is straightforward to show that a subdomain of $\mathcal{U}$
that does not contain
the hypersurface $r=r_0$ is isometric to $D_{EF}$.
For that, it suffices to perform the change of coordinates
$\{\tilde\tau,\rEF\}\to\{\tau,z\}$ defined by
$\tilde\tau(\tau)=\tau$ and $\rEF(z)=r(z)$ with $r(z)$ given above.
We may choose
the positive branch of \eqref{rx}, so that $z$ is restricted by $z>0$.
It is immediate then to obtain from \eqref{metricglp} the line element
\eqref{g:tau_x} restricted to the half plane $z>0$.
As a result, the domain $D_{EF}$ is isometric to the subdomain $\mathcal{U}|_{z>0}$
of $\mathcal{U}$.

On the other hand, the change $\{\tilde \tau,\rEF\}\to\{\text,\rext\}$ defined by
  $\rEF=\rext$ and
  \[
    \text=\tau+\sqrt{2m}\left(2\sqrt{\rext-r_0}+\frac{2m}{\sqrt{2m-r_0}}
      \log\left(\frac{\sqrt{2m-r_0}-\sqrt{\rext-r_0}}{\sqrt{2m-r_0}+\sqrt{\rext-r_0}}\right)\right),
  \]
which is defined for $\rEF=\rext>2m$, brings \eqref{metricglp}
to the form \eqref{deformedschwarzsmetric}. That is,
the subdomain $D_{EF}|_{\rEF>2m}$ of $D_{EF}$ is isometric to $D_S$.
Of course, as a result, the subdomain $\mathcal{U}|_{z>z_s}$, where
$z_s$ is the positive root of $r(z)=2m$, is isometric to $D_S$.
At this point we know that $\mathcal{U}$ covers $D_{EF}$, which in
turn covers $D_S$, but we do not know yet the causal structure of any.

In the next section we focus on the domain $\mathcal{U}$ and find
its global structure in the form of a Penrose diagram. {On the way
we will obtain the diffeomorphisms relating the previously mentioned charts, and thus the causal structure of
$D_{EF}$, $D_S$ and $D_h$.}

\section{Domain $\mathcal{U}$ and global structure of the spacetime solution}
\label{sec:global}
Up to this point, we have a spherically symmetric spacetime $(M,g)$
such that in the chart $\Psi^{\mathcal{U}}_{\tau \xu}=\{\tau,\xu\}$,
defined over some domain $\mathcal{U}\subset M$,
the line element is given by \eqref{g:tau_x}
and the area-radius function $r$, $r(\xu)$, satisfies \eqref{rx}
[or equivalently \eqref{x_r} with $r(-\xu)=r(\xu)$ and $r(0)=r_0$]
with $r_0>0$ and $m>0$.
The domain $\mathcal{U}$ is foliated by the level surfaces (spacelike hypersurfaces) of $\tau$,
and we recall
we take minus the normalized gradient of $\tau$, which reads explicitly 
 $n=-N\nabla\tau=\partial_\tau-\sqrt{2m/r}\partial_\xu$ on $\mathcal{U}$, as the
representative of the future-pointing direction.

In this section we look for the global structure
of $(\mathcal{U},g)$. On the way, we will produce appropriate
coordinate transformations from $(\tau,\xu)$ to null coordinates
so that the metric \eqref{g:tau_x} takes the explicit conformally flat
form on the $(\tau,\xu)$ plane. By doing that in different
regions of the domain $\mathcal{U}$ we will obtain
the previous line elements \eqref{deformedschwarzsmetric}, 
  \eqref{metschwarzshom}, \eqref{g_I_full}, and \eqref{metricglp}
on their corresponding domains.
This will show that $(\mathcal{U},g)$ covers any such static
$D_S$, 
homogeneous $D_h$ 
and $D_H$, 
and Eddington-Finkelstein regions $D_{EF}$. 
The procedure will end by
proving that $(\mathcal{U},g)$ contains exactly one
globally hyperbolic interior homogeneous domain
isometric to a region $D_H$ 
and
two static exterior regions, both isometric to a region $D_S$.
Moreover, we find the maximal analytic extension.
This whole process, along with
the resulting Penrose diagram, is sketched in
Figs.~\ref{diagram:U} and \ref{diagram:maxextension}.
The rest of this section presents the technical details of such construction,
so the reader not interested in those details can move on to the next section,
where we present the physical and geometrical properties of the solution.

Before continuing, let us first define the relevant subsets of $\mathcal{U}$
and the notation we will need in the following.
Using the values of the area-radius function $r$, which we take as a function
defined on $M$, 
we start by defining
three sets on $\mathcal{U}$ as $\exterior:=\{r>2m\}\cap \mathcal{U}$,
$\interior:=\{r<2m\}\cap \mathcal{U}$ and
$\horizon:=\{r=2m\}\cap\mathcal{U}$.
On the other hand,
the condition $r(\xu)>2m$ 
defines for $\xu$ the intervals
$\xu\in (-\infty,-\xu_s)\cup(\xu_s,\infty)$ for all $\tau$,
where $\xu_s$, let us recall, denotes the positive root of $r(\xu_s)=2m$.
We denote the regions that correspond
to $\xu\in (\xu_s,\infty)$ and $\xu\in (-\infty,-\xu_s)$ by the chart $\Psi^{\mathcal{U}}_{\tau \xu}$
as  $\exterior_+$ and $\exterior_-$ respectively,
so that $\exterior=\exterior_+\cup\exterior_-$, which is thus a disconnected region.
Analogously, we define $\interior_+$ and $\interior_-$
to be the domains in $\interior$ with positive
and negative values of $\xu$, respectively.
Observe that
$\horizon$ contains two disconnected sets
that correspond to $\xu=\xu_s$ and $\xu=-\xu_s$ by the chart $\Psi^{\mathcal{U}}_{\tau \xu}$,
which we will denote by $\horizon_+$ and $\horizon_-$ correspondingly.
Finally, we denote by $\mathcal{T}\subset\interior$
the set of points $\{r=r_0\}\cap\mathcal{U}$,
which is connected and mapped to $\xu=0$ by $\Psi^{\mathcal{U}}_{\tau \xu}$.
Let us note that $\interior=\interior_+\cup\mathcal{T}\cup\interior_-$
is a connected domain in $\mathcal{U}$
mapped to the stripe $\xu\in(-\xu_s,\xu_s)$ in $\mathbb{R}^2$.
We will reduce the set of expressions
at the cost of introducing some extra notation as follows.
We define the auxiliary variable $\sigma$ with possible values $+1$ and $-1$,
or $+$ and $-$ when convenient.
Given a value for $\sigma$
we will denote by $D_\sigma:=D|_{\sgn(\xu)=\sigma}$ the restriction of some set $D$
taking the points where the coordinate function $\xu$ satisfies $\sgn(\xu)=\sigma$,
and we will also use $\sigma$ to label their corresponding charts.
Not to overwhelm the notation, when convenient,  
we will denote also by $\exterior$, $\exterior_\sigma$,
$\interior$, etc... the images of the domains on $\mathbb{R}^2$ under any chart.

\subsection{Geodesics}
The radial geodesics of the metric \eqref{g:tau_x}, parametrized as
$
\{\tau(s),\xu(s)\}$ with affine parameter $s$,
are determined by the two equations
\begin{equation}
  \gamma=-\left(1-\frac{2m}{r(\xu)}\right)\left(\frac{d\tau}{ds}\right)^2
  +\left(\frac{d\xu}{ds}\right)^2+2\sqrt{\frac{2m}{r(\xu)}}\frac{d\tau}{ds}\frac{d\xu}{ds},\qquad    \mathcal{E}=    -\bigg(1-\frac{2m}{r(z)}\bigg)\frac{d\tau}{ds} +\sqrt{\frac{2m}{r(z)}}\,\frac{d\xu}{ds},\label{geodesics}
\end{equation}
where $\gamma=0,1,-1$ for null, spacelike and timelike geodesics respectively,
and $\mathcal{E}$ denotes the energy,
which is the conserved quantity associated with the timelike Killing vector field $\partial_\tau$.
The two equations can be combined to obtain
\begin{align}\label{zdot_E}
    \left(\frac{d\xu}{ds}\right)^2 ={\mathcal{E}^2+\gamma\bigg(1-\frac{2m}{r(z)}\bigg)}.
\end{align}

We focus now on the null geodesics, with $\gamma=0$.
If $d\xu/ds\neq 0$ (nonzero energy), then we choose the parameter $s$ so that $d \xu/ds=\signgeod=\pm 1$.
The two possible values of $\signgeod$
produce the ``outgoing'' $k$ ($\signgeod=1$) and ``ingoing'' $l$
($\signgeod=-1$) geodesic vectors explicitly as
\[
  l=\left(1+\sqrt{\frac{2m}{r}}\right)^{-1}\partial_\tau-\partial_\xu,\qquad
  k=\left(1-\sqrt{\frac{2m}{r}}\right)^{-1}\partial_\tau+\partial_\xu.
\]
Observe first that the affine parametrization implies $d\bm l=0$ and $d\bm k=0$,
where we use the boldface to denote 1-forms, so that
$\bm l=l_\mu dx^\mu$ and $\bm k=k_\mu dx^\mu$. 
We must stress that ``outgoing'' and ``ingoing'' are names that only make sense
in the exterior region $\exterior$.
Moreover, $l$, which  is future pointing 
everywhere because  {$g(l,n)=-(1+\sqrt{2m/r})^{-1}<0$},
is ingoing in $\exterior_+$ while it is outgoing in
$\exterior_-$.

If $d\xu/ds=0$ (zero energy, $\mathcal{E}=0$), then $d\tau/ds$ cannot be vanishing,
and thus those geodesic radial curves are $z=\pm z_s$, that is they lie on
$\horizon_-$ and $\horizon_+$,
and the tangent vector is $\partial_\tau|_{r=2m}$, which is null there.
The expression for $l$ defines a smooth
vector field for all values of
$\xu$, and thus at all points in $\mathcal{U}$.
On the other hand,
$k$ is not defined at $z=\pm z_s$, {and,
because} {$g(k,n)=-(1-\sqrt{2m/r})^{-1}$}, it
is future pointing on $\exterior$ and past pointing
on $\interior$.
Observe that $\xu$ is the affine parameter (up to a constant scaling factor)
  of the radial null geodesics on $\mathcal{U}$. This clearly recovers the fact that $r$
is the affine parameter of the radial null geodesics in Schwarzschild.

\subsection{Half-null chart for $\mathcal{U}$}
\begin{figure}
  \centering
  \includegraphics[width=0.5\textwidth]{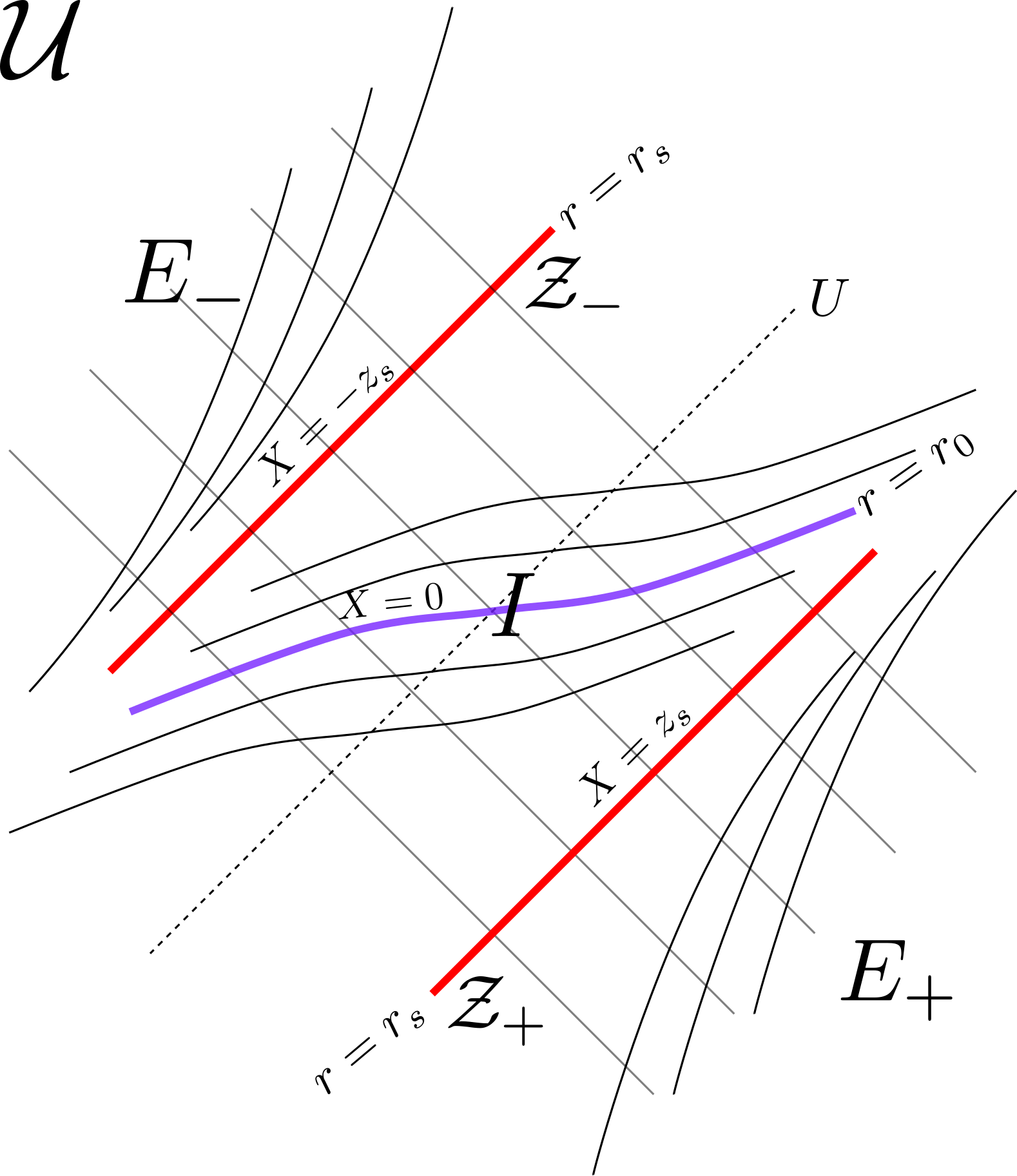}\hfill
  \caption{A diagram of the image of $\mathcal{U}$ under $\Psi^{\mathcal{U}}_{UX}$, omitting
      the angular part. Hypersurfaces of constant $U$ are depicted as parallel
      lines at $-45$ degrees, while those of constant $X$ correspond to the rest of curves
      in black, plus the curve $X=0$ in violet and the lines $X=-\xu_s$ and $X=\xu_s$,
    corresponding to $\horizon_-$ and $\horizon_+$ respectively, in red.
  }
  \label{diagram:UX}
\end{figure}

We can now use $l$ to define all over $\mathcal{U}$ the usual Eddington-Finkelstein
coordinates  by performing the change
$\{\tau,\xu\}\to \{U,X\}$ determined by $X=\xu$ and
\[
\bm{l} 
  =-d\tau-\left(1+\sqrt{\frac{2m}{r(\xu)}}\right)^{-1}d\xu=-dU,
\]
that is,
\[
  \frac{\partial U(\tau,\xu)}{\partial \tau}
  =1,\quad
  \frac{\partial U(\tau,\xu)}{\partial \xu}
  =\left(1+\sqrt{\frac{2m}{r(\xu)}}\right)^{-1}.
\]
Using the relation \eqref{rx} we then find
\begin{equation}
  U(\tau,\xu)=\tau+\sgn(\xu)R_U(r(\xu)),
  \label{U}
\end{equation}
with
\[
  R_U(\arguments)=\int^r_{r_0}\left(\sqrt{1-\frac{r_0}{s}}\left(1+\sqrt{\frac{2m}{s}}\right)\right)^{-1} ds,  
\]
so that this function vanishes at $r_0$, 
 $R_U(r_0)=R_U(r(0))=0$,
and hence provides a function $U(\tau,\xu)$ analytic
on the whole plane
$(\tau,\xu)\in\mathbb{R}^2$. The above integral can be explicitly performed, leading to
\begin{align*}
  R_U(\arguments)=&4m\left(\sqrt{1-\frac{r_0}{2m}}\right)^{-1}
                \log\left(\sqrt{r_0}\frac{\sqrt{r-r_0}+\sqrt{2m-r_0}}
                {\sqrt{2m}\sqrt{r-r_0}+\sqrt{r}\sqrt{2m-r_0}}\right)\\
               &+\left(\sqrt{r}- 2\sqrt{2m}\right)\sqrt{r-r_0}
                +(4m + r_0)\log\left(\sqrt{\frac{r}{r_0}}+\sqrt{\frac{r}{r_0}-1}\right).
\end{align*}
This function is strictly increasing and it can be checked that $\lim_{r\to\infty}R_U(\arguments)/r=1$. 
The change of coordinates $\Phi^{\mathcal{U}}:\{\tau,\xu\}\to \{U,X\}$
thus constructed,
\[
\Phi^{\mathcal{U}}=\left\{  X(\tau,\xu)=\xu,\quad U(\tau,\xu)=
  \tau+\sgn(\xu) R_U(\xu)\right\},
\]
provides a diffeomorphism from
$\mathbb{R}^2$ to $\mathbb{R}^2$.
Therefore the chart $\Psi^\mathcal{U}_{UX}= \{U,X\}$
given by $\Psi^{\mathcal{U}}_{UX}= \Phi^{\mathcal{U}}\circ\Psi^{\mathcal{U}}_{\tau \xu}$ is defined all over $\mathcal{U}$
and the metric \eqref{g:tau_x} reads in that chart
\begin{equation}
  ds^2=-\left(1-\frac{2m}{r(X)}\right)dU^2+2dUdX+r(X)^2d\Omega^2,
  \label{g:V_x}
\end{equation}
where $r$ in this new chart $r(X)$ is just $r(\xu)$, cf. \eqref{x_r}, replacing $\xu$ by $X$.
The hypersurfaces of constant $U$ are obviously null, while
those of constant $X$ are spacelike on $\interior$, timelike on
$\exterior$ and null on $\horizon$. 
We depict in Fig.~\ref{diagram:UX} a qualitative diagram
of the image of $\mathcal{U}$ under the chart
$\Psi^{\mathcal{U}}_{UX}$.

\subsection{Null coordinates}
In order to produce the Penrose diagram for $\mathcal{U}$
we now perform the change to null coordinates
on each disjoint domain $\exterior_+$, $\exterior_-$, and $\interior$ in $\mathcal{U}$.
We start with the domains $\exterior_\sigma$.
For each $\exterior_\sigma$ we construct the corresponding change
$\widehat\Phi_\sigma:\{\tau,\xu\}|_{\exterior_\sigma}\to \{U_\sigma,V_\sigma\}$,
where $\xu|_{\exterior_+}=(\xu_s,\infty)$ and $\xu|_{\exterior_-}=(-\infty,-\xu_s)$,
by taking $U_{\sigma}$ to be the restriction of $U$ \eqref{U} on $\exterior_\sigma$,
i.e., $U_\sigma=U|_{\exterior_\sigma}$,
so that
\[
  U_\sigma(\tau,\xu)=\tau+\sigma R_U(r(\xu))
\]
on each $\exterior_\sigma$, and $V_\sigma$ to satisfy
\[
 \bm{k}|_{\exterior_\sigma} 
  =-d\tau+\left(1-\sqrt{\frac{2m}{r(\xu)}}\right)^{-1}d\xu|_{\exterior_\sigma}
  =-dV_\sigma.
\]
After using \eqref{rx}
we thus have, on each $\exterior_\sigma$,
\[
V_\sigma(\tau,\xu)=\tau -\sigma R^\exterior_V(r(\xu)),
\]
where
\[
  R^\exterior_V(\arguments)=\int\left(\sqrt{1-\frac{r_0}{r}}\left(1-\sqrt{\frac{2m}{r}}\right)\right)^{-1} dr + C_V.\]
A convenient choice of constant $C_V$ leads to the explicit form
\begin{align*}
    R^\exterior_V(\arguments)=&4m\left(\sqrt{1-\frac{r_0}{2m}}\right)^{-1}
                \log\left(\sqrt{r_0}\frac{\sqrt{r-r_0}-\sqrt{2m-r_0}}
                {\sqrt{2m}\sqrt{r-r_0}+\sqrt{r}\sqrt{2m-r_0}}\right)\\
               &+\left(\sqrt{r}+ 2\sqrt{2m}\right)\sqrt{r-r_0}
                +(4m + r_0)\log\left(\sqrt{\frac{r}{r_0}}+\sqrt{\frac{r}{r_0}-1}\right),
\end{align*}
which is analytic in its domain of definition $r\in(2m,\infty)$.
This function is clearly strictly increasing in its domain, and
it  is straightforward to check that $\lim_{r\to\infty}R^\exterior_V(\arguments)/r=1$
and $\lim_{r\to 2m} R^\exterior_V(\arguments)=-\infty$.
Each of the diffeomorphisms $\widehat\Phi_{\sigma}$ thus maps the half
plane $r\in(2m,\infty)$ in $\mathbb{R}^2$ to the whole plane.
As a result, the image of each one of
the charts $\widehat\Psi^\exterior_\sigma=\{U_\sigma,V_\sigma\}$,
given by $\widehat\Psi^\exterior_\sigma=\widehat\Phi_\sigma\circ\Psi^{\mathcal{U}}_{\tau \xu}|_{\exterior_\sigma}$, 
covers $\mathbb{R}^2$. Null infinity $\mathcal{J}$ is reached as
  $|\xu|\to\infty$ ($r\to\infty$)
  and, given the orientations of $l$ and $k$ on each $\exterior_\sigma$,
  $\mathcal{J}^\pm$ are located as follows.
  On $\exterior_+$, $\mathcal{J}^+$
    is reached as $U_+\to +\infty$ and fixed $V_+$,
    and $\mathcal{J}^-$ as $V_+\to -\infty$ with fixed $U_+$.
    On $\exterior_-$, $\mathcal{J}^-$ is reached
    as $U_-\to -\infty$ with fixed $V_-$
    and $\mathcal{J}^+$ as $V_-\to +\infty$ with fixed $U_-$.
  In addition,  as $|\xu|\to\infty$ with fixed $\tau$ (spatial infinity $i^0$) 
we have $(U_+,V_+)\to(\infty,-\infty)$ on $\exterior_+$,
whereas  $(U_-,V_-)\to(-\infty,\infty)$ on $\exterior_-$.
Finally, $\tau\to\pm\infty$ with fixed $\xu$ ($i^\pm$) correspond
to $(U_\sigma,V_\sigma)\to(\pm\infty,\pm\infty)$ respectively.

Let us now analyze the domain $\interior$. To do that we construct the change
$\widehat\Phi^\interior:\{\tau,\xu\}|_{\interior}\to \{U^\interior,V^\interior\}$,
where $\xu|_{\interior}=(-\xu_s,\xu_s)$, 
by taking $U^\interior=U|_{\interior}$,
so that
\[
  U^\interior(\tau,\xu)=\tau+\sgn(\xu) R_U(r(\xu))
\]
on $\interior$,
while we request  $V^\interior$ to satisfy
\[
    -\bm{k}|_I 
  =d\tau-\left(1-\sqrt{\frac{2m}{r(\xu)}}\right)^{-1}d\xu|_{\interior}
  =-dV^\interior,
\]
where we have introduced the minus sign
because the null vector field $k$ on $\interior, $ $k|_\interior$, is past pointing.
We thus have that
\[
  V^\interior(\tau,\xu)=-\tau+\sgn(\xu) R^\interior_V(r(\xu))
\]
on $\interior$ with $R^\interior_V(\arguments)$, which satisfies the same equation
that $R^\exterior_V(\arguments)$, given by
\[
  R^\interior_V(\arguments)=\int^r_{r_0}\left(\sqrt{1-\frac{r_0}{s}}\left(1-\sqrt{\frac{2m}{s}}\right)\right)^{-1} ds,\]
so that $R^{\interior}_V(r_0)=0$
and its domain of definition is the interval $r\in [r_0,2m)$,
ensuring that $V^\interior(\tau,\xu)$ is analytic on $\xu\in(-\xu_s,\xu_s)$.
Its explicit form reads 
\begin{align*}
    R^{\interior}_V(\arguments)=&4m\left(\sqrt{1-\frac{r_0}{2m}}\right)^{-1}
                \log\left(\sqrt{r_0}\frac{\sqrt{2m-r_0}-\sqrt{r-r_0}}
                {\sqrt{2m}\sqrt{r-r_0}+\sqrt{r}\sqrt{2m-r_0}}\right)\\
               &+\left(\sqrt{r}+ 2\sqrt{2m}\right)\sqrt{r-r_0}
                +(4m + r_0)\log\left(\sqrt{\frac{r}{r_0}}+\sqrt{\frac{r}{r_0}-1}\right).
\end{align*}
This function is strictly decreasing with $R^\interior_V(r_0)=0$,
and it is therefore negative on its domain of definition.
Since $\lim_{r\to 2m}R^\interior_V(\arguments)= -\infty$,
for finite values of $\tau$, as one approaches the values $\xu\to \xu_s$
or $\xu\to -\xu_s$ the function $U^\interior$ remains bounded while
$V^\interior$ diverges as $V^\interior\to -\sgn(\xu)\infty$.
Further, boundedness of $R_U(r)$ on $r\in(r_0,2m)$ also implies that
   $U^\interior$ is bounded for finite values of $\tau$,
  and that the limits $\tau\to\pm\infty$ 
  are equivalent to $U^\interior\to\pm\infty$, respectively.
The diffeomorphism $\widehat{\Phi}^\interior$ thus maps
  the stripe $\xu\in(-\xu_s,\xu_s)$ in $\mathbb{R}^2$
  to the whole plane, and therefore
the chart $\widehat\Psi^\interior=\{U^\interior,V^\interior\}$
given by $\widehat\Psi^\interior=\widehat\Phi^\interior\circ\Psi^{\mathcal{U}}_{\tau \xu}|_{\interior}$ maps $\interior$ to the whole $\mathbb{R}^2$.

So far we have  constructed three changes of coordinates from $\{\tau,\xu\}$
that produce three charts, $\widehat\Psi^\exterior_\sigma=\{U_\sigma,V_\sigma\}$ and
$\widehat\Psi^\interior=\{U^\interior,V^\interior\}$, each  mapping
their respective (disjoint) domains to the whole $\mathbb{R}^2$.
If we drop the indexes for $U$ and $V$, the metric
in the respective charts has the form
\begin{equation}
  ds^2=-\left|1-\frac{2m}{r(U,V)}\right|dUdV+r(U,V)^2d\Omega^2,
  \label{g:UV}
\end{equation}
where $r(U,V)$, the form of the area-radius function $r$
  in the corresponding chart, is obtained
implicitly in each case from
\begin{align}
  U_\sigma-V_\sigma=&\sigma(R_U(\arguments)+R^\exterior_V(\arguments))\nonumber\\
  =&\sigma\left\{4m\left(\sqrt{1-\frac{r_0}{2m}}\right)^{-1}
                    \left[\log\left(\frac{r}{2m}-1\right)
                    -2\log\left(\sqrt{\frac{r}{r_0}-1}
                    +\sqrt{\frac{r}{2m}}\sqrt{\frac{2m}{r_0}-1}\right)
                    \right]\right.\nonumber\\
              &\left.+2\sqrt{r}\sqrt{r-r_0}
                +2(4m + r_0)\log\left(\sqrt{\frac{r}{r_0}}+\sqrt{\frac{r}{r_0}-1}\right)\right\},
                \label{V-U_E}
\end{align}
in the exterior $E_\sigma$ domains, and from
\begin{align}
  U^\interior+V^\interior=&\sgn(\xu)(R_U(\arguments)+R^{\interior}_V(\arguments))\nonumber\\
  =&\sgn(\xu)\left\{4m\left(\sqrt{1-\frac{r_0}{2m}}\right)^{-1}
                    \left[\log\left(1-\frac{r}{2m}\right)
                    -2\log\left(\sqrt{\frac{r}{r_0}-1}
                    +\sqrt{\frac{r}{2m}}\sqrt{\frac{2m}{r_0}-1}\right)
                    \right]\right.\nonumber\\
              &\left.+2\sqrt{r}\sqrt{r-r_0}
                +2(4m + r_0)\log\left(\sqrt{\frac{r}{r_0}}+\sqrt{\frac{r}{r_0}-1}\right)\right\},
                \label{V-U_I}
\end{align}
in the interior $I$ domain.

Let us remark that the terms $R_U+R^E_V$ and $R_U+R^I_V$,
each at its corresponding domain of definition,
satisfy the same differential equation
\begin{equation}
  \frac{1}{2}\frac{d(R_U+R_V)}{dr}=\left(1-\frac{r_0}{r}\right)^{\!-1/2}\left(1-\frac{2m}{r}\right)^{\!-1},
  \label{tortoise}
\end{equation}
where the superindices $E$ and $I$ have been removed.
In particular, $R_U+R^\interior_V$ is a strictly decreasing
function of $r$ in its domain $(r_0,2m)$
with $R_U(r_0)+R^\interior_V(r_0)=0$, and
it is thus negative. Therefore, \eqref{V-U_I} implies
$\sgn(U^\interior+V^\interior)=-\sgn(\xu)$.
Conversely, $R_U+R^\exterior_V$
  is strictly increasing on its domain $r\in(2m,\infty)$
  and its image covers the real line.
  
  For convenience, we shall denote
  by $r_*$ the function on $\exterior$ defined by
  \begin{equation}
    r_*:=\frac{1}{2}(R_U+R^\exterior_V),
    \label{def:tortoise}
  \end{equation}
see \eqref{V-U_E},  which reduces to the usual tortoise coordinate in the Schwarzschild
   limit $r_0\to 0$, that is
  \begin{equation}
    \lim_{r_0\to 0} r_*=r+ 2m \log\left(\frac{r}{2m}-1\right).
    \label{limit_tortoise}
  \end{equation}

In the following two subsections we proceed to find a convenient compactification
for each of the three charts.

\subsection{Compactification of the two exterior domains}
Following the standard procedure\footnote{By standard we
  refer to the procedure used, e.g., in \cite{griffiths_podolsky_2009}. In that
  case the conformal mapping procedure only
  extends continuously to the boundary.
  That is enough for our purposes here, although
  alternative differentiable (and analytic) approaches
  used in Schwarzschild, see, e.g., \cite{Hal_ek_2013} and references therein,
  could be translated to the present case in an analogous manner.}
for the two exterior domains $\exterior_\sigma$ we perform the
changes $\Theta_\sigma: \{U_\sigma,V_\sigma\}\to \{u_\sigma,v_\sigma\}$ by defining
\[
\Theta_\sigma=\left\{  u_\sigma=\sigma\arctan\exp\left[\frac{\sigma}{4m}\sqrt{1-\frac{r_0}{2m}}U_\sigma\right],\quad
  v_\sigma=-\sigma\arctan\exp\left[-\frac{\sigma}{4m}\sqrt{1-\frac{r_0}{2m}}V_\sigma\right]\right\}.
\]
The map $\Theta_+$ takes $\mathbb{R}^2$ to the (open) square 
$A_+:=\{(u_+,v_+);u_+\in(0,\pi/2), v_+\in(-\pi/2,0)\}$, while
the image of $\Theta_-$ is  $A_-:=\{(u_-,v_-);u_-\in(-\pi/2,0), v_-\in(0,\pi/2)\}$.
The metric in the charts
$\Psi^\exterior_\sigma=\{u_\sigma,v_\sigma\}=\Theta_\sigma\circ\widehat\Psi^{\exterior}_\sigma$
thus constructed
reads, cf. \eqref{g:UV},
\begin{equation}
  ds^2=\frac{1}{\cos^2u_\sigma\cos^2v_\sigma }\Gamma(r(u_\sigma,v_\sigma))du_\sigma dv_\sigma+r(u_\sigma,v_\sigma)^2d\Omega^2,
  \label{g:uv_ext}
\end{equation}
with
\begin{align}
  \Gamma(\arguments)
  :=&- \frac{32 m^3}{r} \left(1-\frac{r_0}{2m}\right)^{-1}\left(\sqrt{1-\frac{r_0}{r}}
     +\sqrt{1-\frac{r_0}{2m}}\right)^2
     \exp\left[-\sqrt{1-\frac{r_0}{2m}}\sqrt{1-\frac{r_0}{r}}\frac{r}{2m}\right]\nonumber \\
                 &\times  \left(1+\sqrt{1-\frac{r_0}{r}}\right)^{-\sqrt{1-\frac{r_0}{2m}}(2+\frac{r_0}{2m})}
                   \left(\frac{r_0}{r}\right)^{\sqrt{1-\frac{r_0}{2m}}(1+\frac{r_0}{4m})-1},\label{F(r)}
\end{align}
and where
the function $r(u_\sigma,v_\sigma)$ 
is determined implicitly by the relation
\begin{equation}
  \label{eq:r_uv}
  \tan u_\sigma \tan v_\sigma=\left(1-\frac{2m}{r}\right)\frac{32 m^3}{2m-r_0}\frac{1}{\Gamma(\arguments)}=:\Upsilon(\arguments).
\end{equation}
Although we have constructed the function $\Gamma$ on the exterior region
$\exterior$, it is important to note that the definition \eqref{F(r)}
is valid 
for $r\in[r_0,\infty)$, that is, all over $\mathcal{U}$.
In addition, $\Gamma(\arguments)$ is negative,
and satisfies $\Gamma\Upsilon'=8m(1-r_0/2m)^{-1/2}(1-r_0/r)^{-1/2}$.
As a result,
$\Upsilon$ is a strictly decreasing function of $r$ with
$\Upsilon(2m)=0$, and thus negative for $r\in(2m,\infty)$.

To see the explicit analogy with the Schwarzschild case,
it may be useful to remark that the function
$\Gamma(\arguments)$
on $r\in(2m,\infty)$ can be written in terms of the function $r_*$ defined in
\eqref{def:tortoise}
as
\begin{equation*}
  \Gamma(\arguments)=
  -16 m^2\left(1-\frac{2m}{r}\right) \left(1-\frac{r_0}{2m}\right)^{-1}\exp\left[-\frac{1}{2m}\sqrt{1-\frac{r_0}{2m}}r_*(r)\right],
\end{equation*}
from where
\[
  \Upsilon(r)=-\exp\left[\frac{1}{2m}\sqrt{1-\frac{r_0}{2m}}r_*(r)\right], \quad \mbox{ for }
  \quad r\in(2m,\infty).
\]
The limit \eqref{limit_tortoise} leads to 
\begin{equation}
  \lim_{r_0\to 0}\Gamma(\arguments)= -\frac{32m^3}{r}e^{-r/2m},
  \label{limit_gamma}
\end{equation}
yielding the usual Kruskal-Szekeres line element of  Schwarzschild.

Given \eqref{eq:r_uv}, the sets of constant $r$
correspond, one to one, to constant values of the product
$\tan u_\sigma\tan v_\sigma$
and, since $r>2m$ on both $\exterior_\sigma$, this is in agreement with
$\tan u_\sigma\tan v_\sigma<0$ for both $\sigma$ in the ranges of
the images of the charts ${\Psi}^\exterior_\sigma$ given above.
The horizon, located at the
limit $r\to 2m$, corresponds to the segments defined by $\tan u_\sigma\tan v_\sigma=0$,
that is $u_\sigma=0$ and $v_\sigma=0$ on the respective $A_\sigma$.
From the location of the respective limits
  in the charts $\{U_\sigma,V_\sigma\}$,
null infinity
corresponds to $\xu\to-\infty$ on $A_-$,
  and thus to the segments $u_-=-\pi/2$ ($\mathcal{J^-}$)
  and $v_-=\pi/2$  ($\mathcal{J^+}$)  there; and
to $\xu\to+\infty$ on $A_+$, and thus to the segments
$u_+=\pi/2$  ($\mathcal{J^+}$) and $v_+=-\pi/2$  ($\mathcal{J^-}$) there.
Moreover,
$i^0$ is attained on $(u_-,v_-)=(-\pi/2,\pi/2)$ and $(u_+,v_+)=(\pi/2,-\pi/2)$.
On the other hand, 
$i^+$
is to be located on $(u_-,v_-)= (0,\pi/2)$ and  $(u_+,v_+)=(\pi/2,0)$,
while $i^-$ 
is on $(u_-,v_-)= (-\pi/2,0)$ and $(u_+,v_+)=(0,-\pi/2)$ on $A_-$ and $A_+$
respectively.

\begin{figure}
   \centering
  \includegraphics[width=0.45\textwidth]{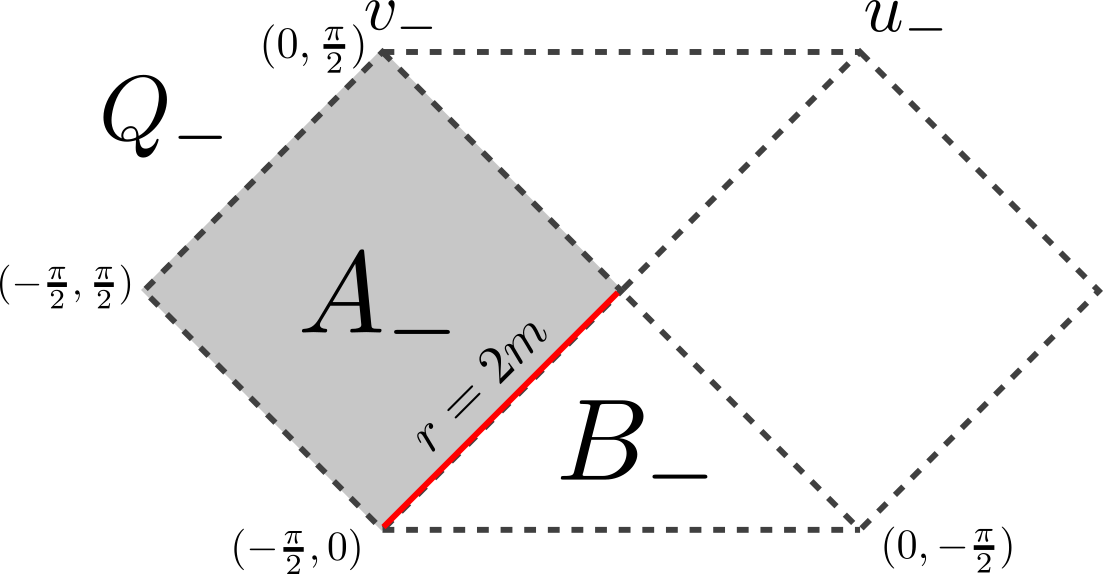}\hfill\includegraphics[width=0.45\textwidth]{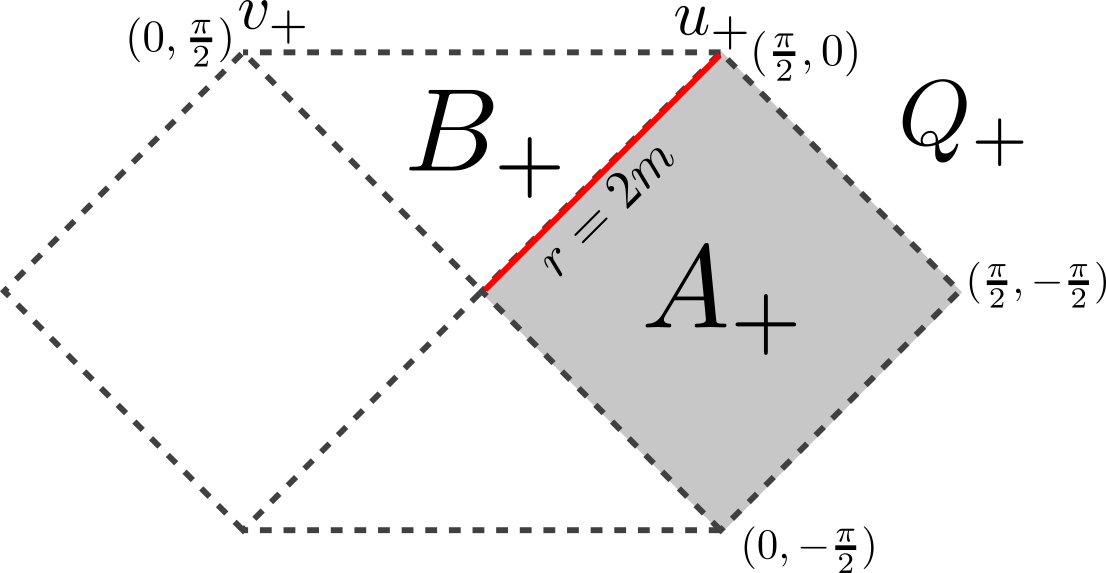}
  \caption{The domains $A_\sigma$ 
    are the images of the exterior domains $\exterior_\sigma$
    through the charts $\Psi^{\exterior}_\sigma$.
    The whole diagrams correspond to $Q_-$ and $Q_+$, the images
    of the extended charts $\Psi^{\mathcal{V}_\sigma}$, so that
    $A_\sigma=\Psi^{\exterior}_\sigma(\exterior_\sigma)
    =\Psi^{\mathcal{V}_\sigma}(\exterior_\sigma)$.}
  \label{diagram:-+}
\end{figure}

The full changes of coordinates
  $\Phi_\sigma:=\Psi^\exterior_\sigma\circ (\Psi^{\mathcal{U}}_{\tau \xu})^{-1}|_{\exterior_\sigma}=\Theta_\sigma\circ\widehat\Phi_\sigma:\{\tau,\xu\}|_{\exterior_\sigma}\to \{u_\sigma,v_\sigma\}$ read
\begin{align}
  \Phi_\sigma=\left\{
  \begin{array}{>{\displaystyle}l}
    u_\sigma=\sigma\arctan\exp\left[\frac{1}{4m}\sqrt{1-\frac{r_0}{2m}} \big(\sigma\tau+R_U(r(\xu))\big)\right],\\\\
    v_\sigma=-\sigma\arctan\exp\left[\frac{1}{4m}\sqrt{1-\frac{r_0}{2m}}\big(-\sigma \tau+R^\exterior_V(r(\xu))\big)\right]
  \end{array}\right\}.\label{Phi_sigma}
\end{align}
  It is direct to check, for consistency, that the maps $\Phi_\sigma$ 
  preserve the time orientation (future is upwards in the diagram)
  by computing the scalar products of
  the vector fields $\partial_{u_\sigma}$ and $\partial_{v_\sigma}$ with {$n$}
  on each respective $A_\sigma$.
  This completes the construction of the Penrose diagram for each of
  the domains $\exterior_\sigma$ as depicted in Fig.~\ref{diagram:-+}
  and shows, in turn, that both $\exterior_\sigma$
    are asymptotically flat and thus the exterior,
    in the sense that the boundaries of the compactified
    domains $A_\sigma$ at infinity have the structure of Minkowski.
    Let us remark (see Sec.~\ref{sec:curvature}) that the Ricci tensor vanishes
  on the boundary at infinity, but not on a neighborhood.

It is now straightforward to check that on each $\exterior_\sigma$ the change
$\{\tau,\xu\}|_{\exterior_\sigma}\to\{\text,\rext\}$, given by $\rext=\rext(\xu)$ and 
$\text=\tau+\frac{\sigma}{2}\left(R_U(r(\xu))-R^\exterior_V(r(\xu))\right)$,
renders the metric \eqref{g:tau_x}, restricted to $r>2m$, in the form \eqref{deformedschwarzsmetric}.
Therefore both $\exterior_\sigma$ are isometric to $D_S$.
This shows that $\mathcal{U}$ covers two exterior regions
(one for each value of $\sigma$) isometric to $D_S$.

\subsection{Compactification of the interior domain}
\begin{figure}
  \centering
  \includegraphics[width=0.3\textwidth]{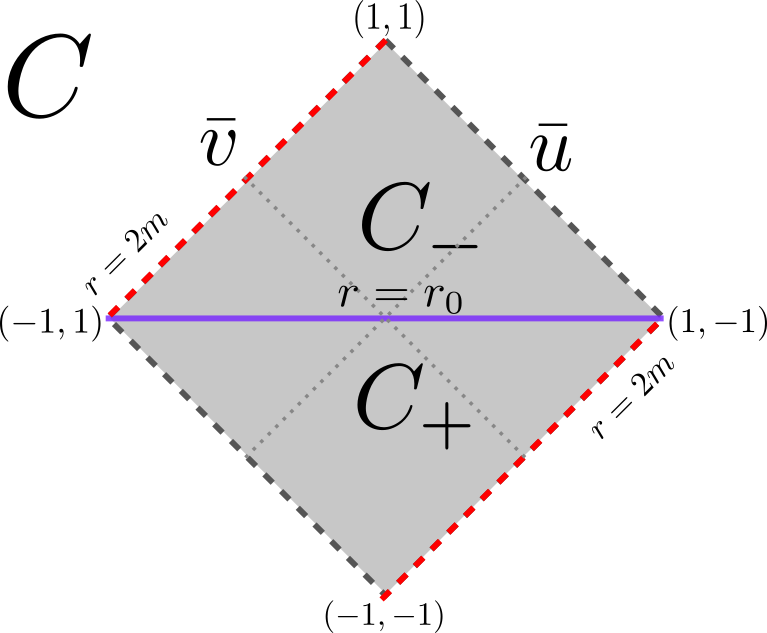}
  \caption{Diagram of the image $C$ of the domain $\interior$ under $\widehat\Psi^\interior$.}
  \label{diagram:C}
\end{figure}

Let us now focus on the domain $\interior$.
We produce a
convenient compact form of the chart $\widehat\Psi^\interior$ for $\interior$ by using the change
$\Theta^\interior:\{U^\interior,V^\interior\}\to \{\uint,\vint\}$ given by
\[
  \Theta^\interior=\left\{
  \uint=\tanh\left(\frac{1}{8m}\sqrt{1-\frac{r_0}{2m}}U^\interior\right),\quad
  \vint=\tanh\left(\frac{1}{8m}\sqrt{1-\frac{r_0}{2m}}V^\interior\right)\right\}.
\]
The map $\Theta^\interior$ takes the whole $\mathbb{R}^2$ plane to the (open) square 
$C:=\{(\uint,\vint);\uint,\vint\in(-1,1)\}$.
The metric \eqref{g:UV} in the chart $\Psi^\interior=\{\uint,\vint\}$ given by
$\Psi^\interior=\Theta^\interior\circ \widehat\Psi^\interior
=\Theta^\interior\circ\widehat\Phi^\interior\circ\Psi^{\mathcal{U}}_{\tau \xu}|_{\interior}$
reads
\begin{equation}
  ds^2=\left(1-\frac{2m}{r(\uint,\vint)}\right)\frac{1}{(1-\uint^2)(1-\vint^2)}
  \frac{128m^3}{2m-r_0}d\uint d\vint
  +r(\uint,\vint)^2d\Omega^2,
  \label{g:uv_int}
\end{equation}
where $r(\uint,\vint)$ satisfies
\begin{align}
  -\left|\frac{\uint+\vint}{1+\uint\vint}\right|
    =\tanh\left[\frac{1}{8m}\sqrt{1-\frac{r_0}{2m}}
      \left(R_U(\arguments)+R^\interior_V(\arguments)\right)\right].
    \label{uv_int_R}
\end{align}
A more explicit expression can be obtained by using
\eqref{V-U_I}.
Since $R_U(r_0)=R^\interior_V(r_0)=0$
the curve $r=r_0$ corresponds to the line $\uint+\vint=0$.
Moreover, since $\sgn(U^\interior+V^\interior)=\sgn(\uint+\vint)$,
  and hence $\sgn(\uint+\vint)=-\sgn(\xu)$,
the curves of constant $r\in(r_0,2m)$ correspond now to $\uint\vint<1$
and thus to two curves of constant $(\uint+\vint)/(1+\uint\vint)$
that go from
$(\uint,\vint)=(-1,1)$ to $(1,-1)$, one
through positive values of $\uint+\vint$ (for $\sgn(\xu)=-1$)
and the other through negative values of $\uint+\vint$ (for $\sgn(\xu)=1$).
As a result, the images $C_\sigma$
of the restrictions of
$\Psi^\interior|_{\interior_\sigma}:\interior_\sigma\to C_\sigma$
are given by $C_+=\{C;\uint+\vint<0\}$
and $C_-=\{C;\uint+\vint>0\}$.
On the one hand, since  as $r\to 2m$ 
the function $R_U$ remains bounded and
$R^\interior_V\to -\infty$,
for finite values of $\tau$ we have that $\uint$ remains bounded
and  $\vint\to -\sgn(\xu)$. Therefore, the curves in $\interior$ approaching
$\horizon_+$ must approach $\vint\to -1$ in the image of the
chart $\Psi^\interior$, whereas
those approaching $\horizon_-$ must tend to $\vint\to 1$.
In other words, $\horizon_-$ and $\horizon_+$ are part of the boundary
of $\interior$, and correspond to the part $\vint=+1$ and $\vint=-1$
[with $\uint\in(-1,1)$] of the boundary of $C$, respectively.
On the other hand, recalling the behavior of $\tau$ and $U^\interior$,
  we have that the limits $\tau\to\pm\infty$ are located at $\uint=\pm 1$.
In Fig.~\ref{diagram:C} we depict
the image {$C$} of the domain $\interior$
through the chart $\Psi^\interior$.

The full change of coordinates
  $\Phi^I:=\Psi^\interior\circ (\Psi^{\mathcal{U}}_{\tau \xu})^{-1}|_{\interior}=\Theta^\interior\circ\widehat\Phi^\interior:\{\tau,\xu\}|_{\interior}\to \{\uint,\vint\}$, reads
\begin{align}
  \Phi^\interior=\left\{
  \begin{array}{>{\displaystyle}l}
    \uint=\tanh\left[\frac{1}{8m}\sqrt{1-\frac{r_0}{2m}}
    \big(\tau+\sgn(\xu)R_U(r(\xu))\big)\right],\\\\
    \vint=\tanh\left[\frac{1}{8m}\sqrt{1-\frac{r_0}{2m}}\big(-\tau+\sgn(\xu)R^\interior_V(r(\xu))\big)\right]
  \end{array}\right\}.\label{Phi_int}
\end{align}
As in the exterior cases, it is straightforward to check that the vector fields
$\partial_{\uint}$ and $\partial_{\vint}$ on $\interior$
are future pointing because they have
negative scalar product with {$n$}, that is,
$\Phi^\interior$ 
preserves the time orientation.

Now, the change $\{\tau,\xu\}|_{\interior}\to \{\timeI,\xuI\}$ given
  by $\tau=\xuI-\frac{\sgn(z)}{2}\left(R_U(\rint(\timeI))-R^\interior_V(\rint(\timeI))\right)$
  and $\rint(\timeI)=r(\xu)$, that implies $\sgn(\xu)=\sgn(\cos\timeI)$ and
  reads, in a more explicit form,
  \begin{align*}
    \tau=&\xuI-4m\left(1-\frac{r_0}{2m}\right)^{-\frac{1}{2}}\,\mbox{artanh}\left[\sqrt{\frac{\rint(\timeI)}{2m}}\cos\timeI\right]
           +4 m\sqrt{1-\frac{r_0}{2m}}\sqrt{\frac{\rint(\timeI)}{2m}}\cos\timeI,\\
    \xu=&r_0\,\mbox{artanh}\left[\sqrt{1-\frac{r_0}{2m}}\cos \timeI\right]
          +\frac{1}{2m}\left(1-\frac{r_0}{2m}\right)^{-\frac{1}{2}}\rint(\timeI)\cos\timeI,
  \end{align*}
  with $\rint(\timeI)$ given by relation \eqref{r_T},
  is a diffeomorphism between the stripe $\xu\in(-\xu_s,\xu_s)$ on $\mathbb{R}^2$
  and the stripe $\timeI\in(0,\pi)$ on $\mathbb{R}^2$ that
  takes the line element \eqref{g:tau_x} to the form \eqref{g_I_full}.
  This shows that $\interior\subset\mathcal{U}$ is isometric
  to the region $D_H$. Observe that $\horizon_+$ ($z=z_s$) is recovered
  for $\timeI=0$;  $\horizon_-$ ($z=-z_s$) for $\timeI=\pi$; and
  $\mathcal{T}$ ($z=0$) for $\timeI=\pi/2$.

  Finally, on each $\interior_\sigma$ the change
  $\{\tau,\xu\}|_{\interior_\sigma}\to \{T,Y\}$, given by $T=r(\xu)$ and
  $
  Y=\tau+\frac{\sigma}{2}\left(R_U(r(\xu))-R^\interior_V(r(\xu))\right)
  $
  brings \eqref{g:tau_x}
  to the form \eqref{metschwarzshom}.
  Therefore $\interior$, and thus $\mathcal{U}$,
  covers two interior regions isometric to $D_h$.
   Note that the change for $\sigma=1$ preserves the time orientation, whereas that for $\sigma=-1$ does not.

  The null radial geodesics clearly spend a finite amount of affine
  parameter to cross the interior domain $\interior$.
  The timelike radial curves, as it will be shown
  in Sec.~\ref{sec:properties} by a direct calculation, spend also
  a finite proper time going from $\xu=\xu_s$ to $\xu=-\xu_s$.

\subsection{Full Penrose diagram and maximal analytic extension}\label{sec:maxextension}

To sum up, starting from the original chart $\Psi^{\mathcal{U}}_{\tau \xu}$
with line element \eqref{g:tau_x},
in the previous subsections we have
constructed charts for the disjoint domains $\exterior_\sigma$
and $\interior$ with images on $\mathbb{R}^2$ producing
corresponding explicit Penrose diagrams.
We have also shown how both $\exterior_\sigma$ are isometric to the previous domain $D_S$
and each one of the $\interior_\sigma\subset \interior$ are isometric to $D_h$.
In order to, first, construct the full diagram for $\mathcal{U}$
and, second, obtain its maximal analytic extension,
we need to extend the charts of the domains $\exterior_\sigma$ within $\mathcal{U}$.

Recalling that $\Gamma(r)$ \eqref{F(r)} is defined all over $r\in[r_0,\infty)$,
and given the properties of $\Upsilon(\arguments)$,
equation \eqref{eq:r_uv} has a solution
for $r(u,v)$ for all pairs of values $(u,v)$ in $\mathbb{R}^2$
for which $r\geq r_0$.
In particular, since $\Gamma(r_0)=-32m^3/r_0$ then $\Upsilon(r_0)=1$ and
we have from \eqref{eq:r_uv} that
$\tan u_\sigma \tan v_\sigma|_{r=r_0}=1$. Therefore
$(u_\sigma+v_\sigma)|_{r=r_0}=\pm \pi/2$.
Also, we have $r(u_\sigma,v_\sigma)=2m$
 if and only if $\tan u_\sigma\tan v_\sigma=0$.
As a result,
we can extend the domains of the charts  $\Psi^\exterior_\sigma$
across the sets where $r=2m$
to two domains $\mathcal{V}_\sigma\subset M$ as the preimages of
two charts $\Psi^{\mathcal{V}_\sigma}:\mathcal{V}_\sigma(\subset M)\to
Q_\sigma(\subset\mathbb{R}^2)$  
with $Q_\sigma=\{(u_\sigma,v_\sigma);u_\sigma,v_\sigma\in(-\pi/2,\pi/2),|u_\sigma+v_\sigma|< \pi/2\}$,
with boundaries $|u_\sigma+v_\sigma|=\pi/2$ located at points where $r=r_0$.
We will also need the convenient 
definitions of the open triangles
$B_\sigma\subset Q_\sigma$ as
$B_+:=\{(u_+,v_+);u_+,v_+\in(0,\pi/2), u_++v_+<\pi/2\}$ and
$B_-:=\{(u_-,v_-);u_-,v_-\in(-\pi/2,0), u_-+v_->-\pi/2\}$.

In order to construct the full diagram for $\mathcal{U}$ 
we need to 
request that the relevant
parts of the extensions of $\exterior_\sigma$ correspond to part of the
interior domain within our original $\mathcal{U}$, that is, 
$\mathcal{V}_\sigma\cap\interior=\interior_\sigma$,
so that the conformal diagram for $\mathcal{U}$ consists of the diagram of $\interior$
patched to the relevant part of the conformal diagram of
$\mathcal{V}_+$ and $\mathcal{V}_-$.
This is done constructively by producing the diffeomorphisms
$\trianglephi_\sigma:C_\sigma\to B_\sigma$
as
\[
  \trianglephi_\sigma=
  \left\{u_\sigma=\sigma\arctan\left(\frac{1+\uint_\sigma}{1-\uint_\sigma}\right)^{\sigma},
  v_\sigma=\sigma\arctan\left(\frac{1+\vint_\sigma}{1-\vint_\sigma}\right)^{\sigma}
  \right\},
\]
where we use $\uint_\sigma=\uint|_{\interior_\sigma}$ and $\vint_\sigma=\vint|_{\interior_\sigma}$.
Now, we only need to
build the charts $\Psi^{\mathcal{V}_\sigma}$
by demanding that for any point $p\in \interior_\sigma$
we have $\Psi^{\mathcal{V}_\sigma}(p)=\trianglephi_\sigma\circ \Psi^\interior(p)$.
By construction, the change of coordinates between the charts
$\Psi^\interior$ and $\Psi^{\mathcal{V}_\sigma}$
restricted to $\interior_\sigma$ are given by
$\Lambda_\sigma$, and the image of each $\interior_\sigma$
is indeed their respective $B_\sigma$.

The horizons $\horizon_+$ and $\horizon_-$ are clearly
  included in $\mathcal{V}_\sigma$ by continuity.
  In fact it is easy to see that the charts $\Psi^\interior$
  can be extended to maps that include $\horizon_+$ and $\horizon_-$
  so that these sets are, respectively, mapped to $\vint=-1$ and $\vint=1$
  at the boundary of $C$. Those points can be mapped, in turn,
  to the corresponding boundaries of $B_\sigma$ by
  adding the relations  $\{v_\sigma=\sigma\pi/2 \Leftrightarrow \vint_\sigma=-\sigma\}$
 to $\Lambda_\sigma$ for each value of $\sigma$.

\begin{figure}
  \centering
  \includegraphics[width=0.9\textwidth]{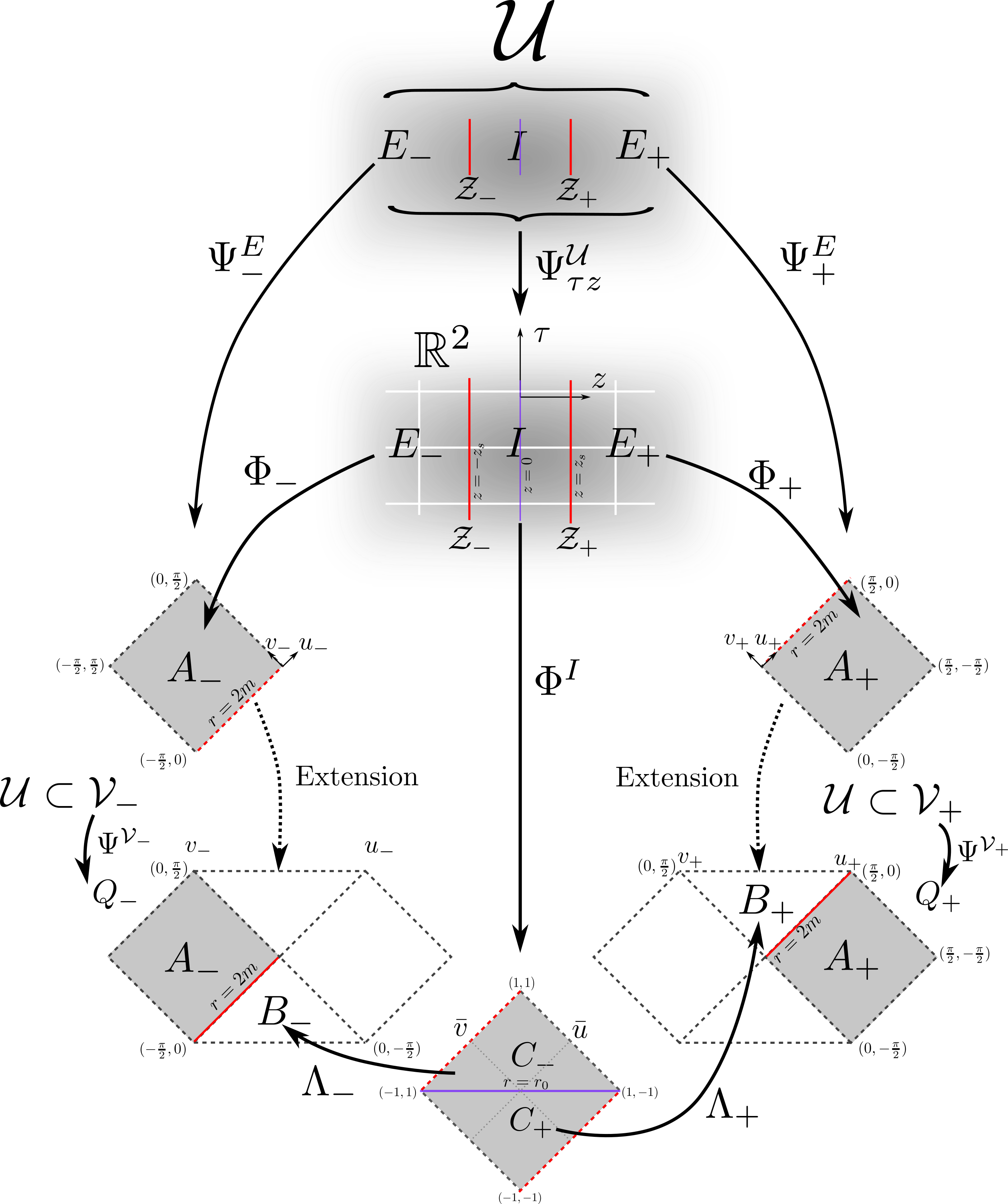}
  \caption{The construction of the Penrose diagram of $(\mathcal{U},g)$.}
  \label{diagram:U}
\end{figure}

\begin{figure}
  \centering
  \includegraphics[width=0.8\textwidth]{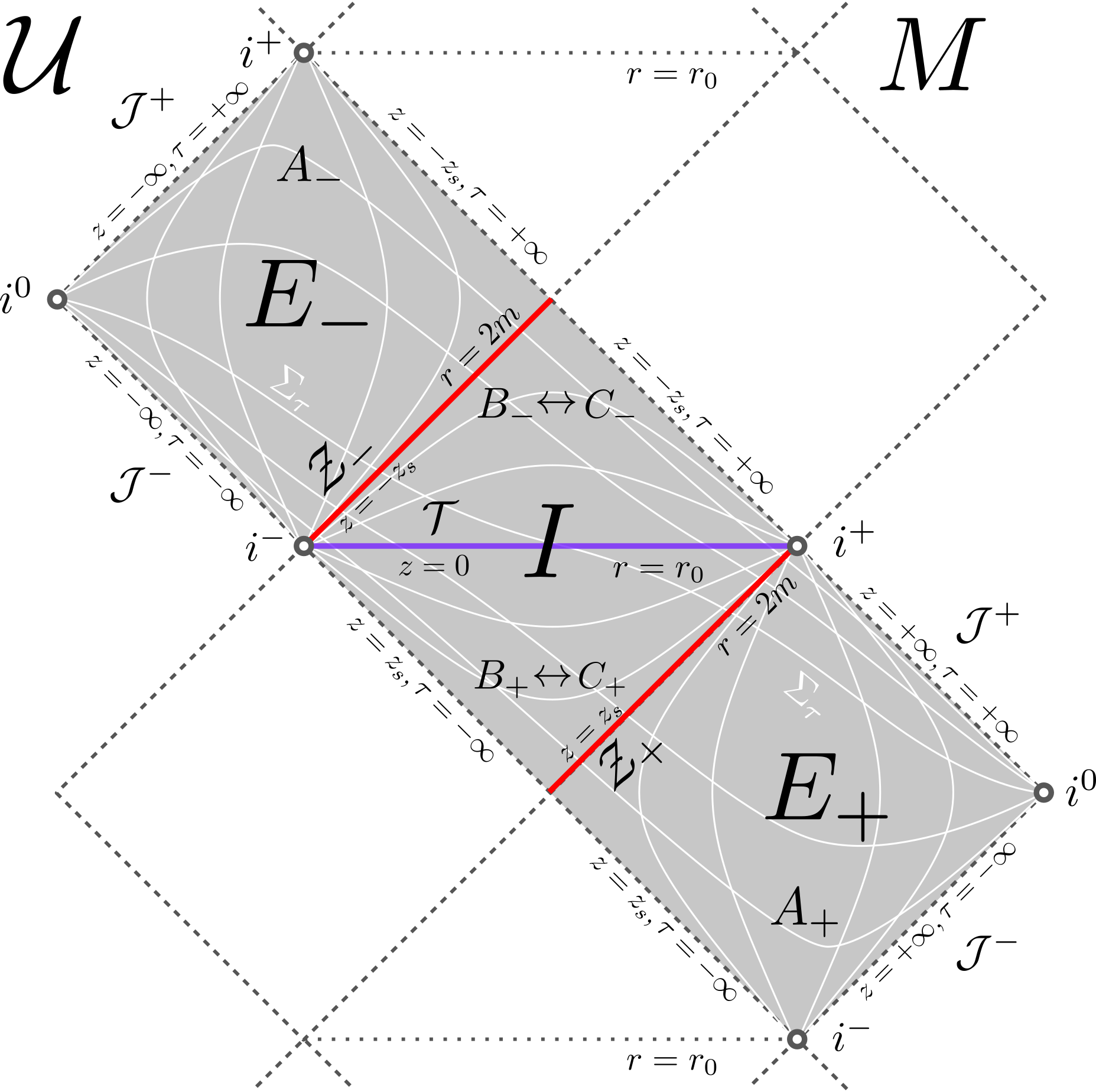}
  \caption{Penrose diagram of $(\mathcal{U},g)$ (shaded),
      and its maximal extension $(M,g)$ (outlined). The curves of constant
      $\xu$ and constant $\tau$ (denoting also $\Sigma_\tau$) are drawn
      in white, and correspond to the white lines depicted on $\mathbb{R}^2$
      in Fig.~\ref{diagram:U}.}
  \label{diagram:maxextension}
\end{figure}

With the above construction, sketched in Fig.~\ref{diagram:U},
we have shown how the initial domain $\mathcal{U}$
can be conformally mapped to the Penrose diagram depicted in Fig.~\ref{diagram:maxextension}, and how it covers all the domains presented in the previous section.
The procedure has provided in a direct manner the causal diagrams of the regions
$D_h$, $D_H$ and $D_S$.
As for the region $D_{EF}$, it only remains to recall
that $D_{EF}$ is isometric to the subdomain $\mathcal{U}|_{z>0}\subset \mathcal{U}$
that, looking at the diagram for $\mathcal{U}$, corresponds to the
same trapezoidal diagram as the Eddington-Finkelstein chart provides for Schwarzschild.

Finally, we can now use the Kruskal-Szekeres type extensions $Q_\sigma$ 
to analytically extend $\mathcal{U}$ to the two domains $\mathcal{V}_\sigma$.
Making use of the usual periodic construction we can build up the maximal
analytic extension, that we denote by $M$ for simplicity, see Fig.~\ref{diagram:maxextension}, so that
any of the domains $\mathcal{V}_\sigma$ 
constitutes the \emph{fundamental domain} of $M$.
Moreover, given that the boundary of the diagram
is exclusively given by sets of the type
$i^0$, $i^\pm$ and $\mathcal{J}^\pm$, we can infer
from the Penrose diagram that $M$ is geodesically complete.

\section{Geometrical and physical properties of the solution}
\label{sec:properties}

In this section we analyze some relevant properties of the spacetime solution $(M,g)$. We start with a review of its causal structure (Sec.~\ref{sec:trapped}), study the curvature (Sec. \ref{sec:curvature}), compare different geometrical masses (Sec. \ref{sec:mass}) and compute the effects on a scalar field propagating on the exterior region (Sec. \ref{sec:qnm}). Finally, we will explicitly obtain the GR limit of the model (Sec. \ref{sec:limits}).

\subsection{Trapped and antitrapped regions}\label{sec:trapped}

The key difference between the manifold $M$ and that of Schwarzschild
is that the function $r$ attains a minimum $r_0>0$ at a certain
spacelike hypersurface $\mathcal{T}$, and that prevents the
singular behavior there.
The hypersurface $\mathcal{T}$, given by $r=r_0$, is in fact
minimal, that is, it has vanishing extrinsic curvature. 
Moreover, the mean curvature vector
of the spheres (surfaces with constant $r$ and $\tau$),
given by\footnote{We take the convention
used, e.g., in \cite{mars_2003}.} $H=(2/r)\nabla r$, reads explicitly
\[
  H=\sgn(\xu)\frac{2}{r^2}\sqrt{1-\frac{r_0}{r}}\left(\sqrt{2mr}\partial_\tau-(2m-r)\partial_\xu\right).
\]
Therefore the scalar products
\[
  g(H,H)=\frac{4}{r^4}(r-r_0)(r-2m),\qquad g(H,n)=-2\frac{\sqrt{2m}}{r^2}\sgn(\xu)\sqrt{r-r_0},
\]
establish that $H$ is spacelike for $r>2m$, this is at $\exterior$,
null at the horizon $\horizon$ ($r=2m$), and timelike for $r_0<r<2m$, where
it is future pointing for $z>0$, i.e., on $\interior_+$, and past pointing for $z<0$,
i.e., on $\interior_-$. This means that the spheres are nontrapped
in the exterior regions, marginally trapped at the horizon, and trapped
to the future in $\interior_+$ (the black-hole region)
and trapped to the past (antitrapped) in $\interior_-$ (the white-hole region),
as expected (see Fig.~\ref{diagram:maxextension}).
Remarkably, the hypersurface $r=r_0$ is characterized by the vanishing of $H$,
meaning that the transition hypersurface $\mathcal{T}$ is foliated by
totally geodesic surfaces of the same area $4\pi r_0^2$.
More explicitly, $\mathcal{T}$ is $\mathbb{R}\times \mathbb{S}^2$
with metric $(2m/r_0 -1)d\tau^2+r_0^2 d\Omega^2$.
Therefore, \emph{$r_0$ encodes the minimal area of the orbits of
the $SO(3)$ group that acts on $M$ by isometry.}

\subsubsection*{Proper time in crossing the homogeneous region}

Let us recall that $\xu$ is, up to a constant multiplicative factor,
  the affine parameter of the radial null geodesics,
  and therefore, they traverse the interior homogeneous region $\interior$
  in a finite amount $2\xu_s$
  of the affine parameter.
For completeness, and to provide another check of the regular behavior of the spacetime
inside the horizon,
we compute the time spent by a radial free falling particle
to cross $\interior$.
For particles initially at rest at infinity, so that $\gamma=-1$ and $\mathcal{E}=-1$,
\eqref{zdot_E} yields
\[
  \frac{d\xu}{ds} =- \sqrt{\frac{2m}{r(z)}},
\]
where we have chosen the sign so that $\xu$ decreases as the affine
parameter $s$ increases. 
As a result, the proper time $\Delta \propertime$
spent to cross from $r=2m$ at $\xu=\xu_s$ to
$r=2m$ at $\xu=-\xu_s$ is
\[
  \Delta \propertime =\int_{\xu_s}^{-\xu_s} ds
  =\int_{-\xu_s}^{\xu_s} \sqrt{\frac{r(\xu)}{2m}}d\xu
  =2\int_{0}^{\xu_s} \sqrt{\frac{r(\xu)}{2m}}d\xu
  =2\int_{r_0}^{2m} \sqrt{\frac{r}{2m}}\left(1-\frac{r_0}{r}\right)^{-1/2}dr= \frac{8}{3}(m+r_0)\sqrt{1-\frac{r_0}{2m}},
\]
after using $r(-z)=r(z)$ in the third equality and \eqref{rx} in the fourth.
This result can be compared with the time spent by an analogous
geodesic in Schwarzschild, that takes $4m/3$ to fall in the singularity. The GR limit of this result, $\Delta\propertime|_{r_0=0}=8m/3$, corresponds to twice the time spent by a radial free-falling observer (initially at rest at infinity) to go from the horizon to the singularity.

\subsection{Curvature}\label{sec:curvature}
Given the trapped nature of the black-hole region $\interior_+$,
one expects the Einstein tensor $G^a{}_{b}$
to have eigenvalues that become negative there
as to avoid the singularity.
Indeed, the Einstein tensor on $\mathcal{U}$, which in the chart $\{\tau,\xu\}$
has the form
\[
  G_{\mu\nu}dx^\mu dx^\nu=r_0\frac{2m}{r^5}(r-2m)d\tau^2
  -2r_0\left(\frac{2m}{r^3}\right)^{\frac{3}{2}}d\tau d\xu
  -\frac{r_0}{r^4}(r+2m)d\xu^2+r_0\frac{r-m}{2 r^2}d\Omega^2,
\]
has two eigenvalues given by
$\mu_1=-r_0/r^3$, $\mu_2=-2mr_0/r^4$ at the Lorentzian part $\{\tau,\xu\}$,
plus $\mu_\varphi=r_0(r-m)/(2r^4)$ at the angular part.
The eigendirections
relative to $\mu_1$ and $\mu_2$ are given by
$v_1=\sqrt{2m/r}\partial_\tau+(1-2m/r)\partial_x$ and
$v_2=\partial_\tau$, with moduli $g(v_1,v_1)=1-2m/r=-g(v_2,v_2)$.
Therefore, interpreting ${G^a}_b$ as an effective energy-momentum tensor on the exterior region $\exterior$
would correspond to a positive ``effective energy'' density $\tilde\rho^\exterior=-\mu_2=2mr_0/r^4$
and a negative effective radial pressure
$\tilde p^\exterior_r=\mu_1=-r_0/r^3$, while on the interior region $\interior$
the effective energy density would turn to be
$\tilde\rho^I=-\mu_1=r_0/r^3$ and the effective radial pressure
$\tilde p^I_r=\mu_2=-2mr_0/r^4$.
As a result, since $\mu_1-\mu_2$ is negative on $\exterior$
and $\mu_2-\mu_1$ is negative on $\interior$,
none of the 
``effective energy'' conditions is satisfied anywhere outside
the horizon.
Let us stress that $(M,g)$ \emph{solves
the vacuum equations} and therefore satisfies trivially
all the \emph{physical} energy conditions by construction.

It is remarkable, however, that on the horizon
the Einstein tensor has a double null eigendirection along $\partial_\tau|_{r=2m}$,
so that
the effective energy density and the radial pressure are equal up to a sign,
that is 
$(\mu_1-\mu_2)|_{r=2m}=0$. Moreover, the effective energy density
and the effective angular pressure on $r=2m$ are positive, and satisfy
$\tilde\rho|_{r=2m}=-\mu_{1}|_{r=2m}=-\mu_{2}|_{r=2m}=4\mu_\varphi|_{r=2m}=r_0/(2m)^3$.
Therefore, all the ``effective energy'' conditions are fulfilled on the horizon.

Furthermore, the four eigenvalues of the Einstein tensor
decay (at least) as $O(r^{-3})$.
Hence all the ``effective energy'' conditions are fulfilled
also at infinity, although not in a neighborhood.

The Ricci scalar reads
\[
  R=\frac{3mr_0}{r^4},
\]
so that it is everywhere positive and attains its maximum
value on the transition hypersurface $\mathcal{T}$, given by $R|_{\mathcal{T}}=3m/r_0^3$.
To finish with the components of the curvature,
the only nonvanishing component of the Weyl tensor is
\[
  \Psi_2=-\frac{m}{r^3}+\frac{r_0}{4r^4}(5m-r),
\]
in the usual null frame adapted to the spherical symmetry. 
For completeness, and to compare with Schwarzschild,
let us include the Kretschmann scalar,
\[
    R_{abcd}R^{abcd}=\frac{48 m^2+24 m r_0+6 r_0^2}{r^6}-\frac{120 m^2 r_0 +32 m r_0^2}{r^7}+\frac{81 m^2 r_0^2}{r^8},
\]
which is always positive and attains its maximum value
${9m^2}/{(2r_0^{6})}$ on $\mathcal{T}$.

\subsection{Mass}\label{sec:mass}
We refer to the constant of motion $m$ as the ``mass''
mainly because \eqref{eq.masspol} is a constant of motion and coincides with
the expression for the
Schwarzschild mass in phase-space variables 
upon the canonical transformation \eqref{cantransf} (see \cite{Alonso-Bardaji:2021tvy}).
In addition, the relation between the radius of the horizon $\horizon$ and
$m$ is $r|_{\horizon}=2m$. 
However, the meaning of $m$, as well as the discussion on the mass, 
needs a
more detailed analysis. Clearly, the addition to $m$ of any function
of $r_0$ provides a constant of motion as well.
To get a proper understanding of these parameters, 
we devote this subsection to present the expressions
of some usual geometrical definitions of mass and energy applied to this solution.

It is important first to stress that the Hawking mass $M_{H}$
and the Komar mass $M_K$
do not need to coincide, both locally and at infinity, because $g$ does
not solve the Einstein equations \cite{Beig_masses_78} (see also \cite{szabados_mass}). 
A direct calculation shows that the Komar mass on any sphere $S$ of
constant radius $\rext=r|_{\exterior}>2m$ and $\tilde{t}$ in any exterior (static) region
depends on $\rext$, and reads \cite{wald2010general}
\begin{equation}
  M_K(\rext):=-\frac{1}{8\pi}\int_S\bm{\epsilon}_{abcd}\nabla^c(\partial_{\tilde{t}})^d
  =m\sqrt{1-\frac{r_0}{\rext}},
\end{equation}
where $\bm{\epsilon}$ is the volume element of $(M,g)$.
The dependence of the Komar mass on $\rext$ comes from the fact 
that the Ricci tensor on $(M,g)$, in fact its
part orthogonal to the spherical orbits, is not zero.
The limit of the Komar mass at infinity is precisely half of the radius of the horizon, that is, it coincides with the constant of motion \eqref{eq.masspol},
\begin{equation}
   m=\lim_{r\to\infty}M_{K}(r).
\end{equation}
On the other hand,
the Hawking mass computed on  any sphere of radius $r\in[r_0,\infty)$
(equivalently, the Misner-Sharp mass)
also depends on $r$ and is given by
\begin{equation}
  M_{H}(r)=\frac{r}{2}(1-g(\nabla r,\nabla r))=
  \frac{r}{2}\left(1-\frac{r^2}{4}g(H,H)\right)=m+\frac{r_0}{2}-\frac{m r_0}{r}.
  \label{M_ms}
\end{equation}
This is always positive and, in particular, coincides with $m$ at the horizon (and only there).
The existence of a nonvanishing Ricci tensor
affects the Komar and Hawking masses in a different manner.

To further discuss other definitions of energy we consider next
the computation of the ADM mass in two slicings:
the hypersurfaces $\Sigma_{\text}$
for constant values of $\text$ on any exterior domain $\exterior_\sigma$,
and the hypersurfaces $\Sigma_\tau$ on $\mathcal{U}$
defined as those of constant $\tau$.

The hypersurfaces $\Sigma_{\text}$ are $(2m,\infty)\times\mathbb{S}^2$
with metric $d\sigma_{\text}^2=(1-r_0/\rext)^{-1}(1-2m/\rext)^{-1}d\rext^2+\rext^2d\Omega^2$
and vanishing extrinsic curvature. In the Penrose diagram,
each $\Sigma_{\text}$ reaches from the bifurcation of the horizon
to spatial infinity $i^0$.
The hypersurfaces $\Sigma_{\text}$ satisfy the suitable fall-off conditions
for asymptotic flatness, and the Ricci scalar, given by ${}^{(3)}\!R_{\text}=4mr_0/r^4$,
is integrable.
Therefore the ADM mass is a geometric invariant \cite{Bartnik_1986},
and it corresponds to the limit of the Hawking mass at $i^0$ \cite{szabados_mass},
\begin{equation}\label{eq.adm}
  M^{\text}_{ADM}=\lim_{r\to\infty}M_{H}(r) =m+\frac{r_0}{2}.
\end{equation}
This result, as well as the expression \eqref{M_ms}, are to be expected,
because the hypersurfaces $\Sigma_{\text}$
can be embedded, with vanishing extrinsic curvature,
in Reissner-Nordstr\"om spacetime with mass $M_{RN}=m+r_0/2$ and charge $Q^2=2m r_0$
(observe $M_{RN}^2-Q^2=(m-r_0/2)^2$ is positive because $0<r_0<2m$).
The fact that the asymptotic properties of the Ricci tensor
  allow the computation of the ADM mass on some slicing with the proper
  fall-off conditions, so that it is indeed a proper invariant quantity,
  is another interesting property of the solution. Let us stress that this
property ought not to be taken for granted, see, e.g., \cite{New_Masses}.

Regarding the hypersurfaces $\Sigma_\tau$, they are
$\mathbb{R}\times\mathbb{S}^2$,
now reaching from one to another component of $i^0$
crossing the hypersurface $\mathcal{T}$, as depicted in Fig.~\ref{diagram:maxextension}.
Their metric reads $d\sigma_\tau^2=d\xu^2+r(\xu)^2d\Omega^2$,
the Ricci scalar ${}^{(3)}\!R_{\tau}$ vanishes (but not the whole Ricci tensor),
and the extrinsic curvature $K_{ab}$ is given by
\[
  K_{\mu\nu}dx^\mu dx^\nu=\sgn(\xu)\sqrt{r-r_0}\sqrt{2m}\left(\frac{1}{2r^2}d\xu^2-d\Omega^2\right).
\]
While the metric goes as $O(1/r)$ at infinity, $K_{ab}$ goes as $O(r^{-3/2})$,
and therefore $\Sigma_{\tau}$ do not satisfy the fall-off conditions
for asymptotic flatness  \cite{szabados_mass}.
As a consequence, the ADM mass does not correspond
necessarily to the limit of the Hawking mass.
In fact, in the Schwarzschild limit
the hypersurfaces $\Sigma_{\tau}$ are defined by
constant $\tau$ in the Painlev\'e-Gullstrand
coordinates, go from (one component of) $i^0$ to the component of $i^+$
that belongs to the other asymptotic end, and are known to be flat (see, e.g., \cite{Alicia_2010}) and therefore have vanishing ADM mass. 
In the present case a direct calculation
of the ADM mass on any $\Sigma_\tau$ provides
\[
  M^{\tau}_{ADM}=\frac{r_0}{2}.
\]
This recovers the result in the GR limit and provides a characterization
of the parameter $r_0$.

We can also consider the Geroch energy.
Let us briefly introduce its motivation, as explained in \cite{szabados_mass}.
The Hawking mass on some
closed codimension two surface $\mathcal{S}$ in $M$
depends, by construction, only on the module of the mean curvature
vector $H$.
Now, given some $\Sigma$,
the Hawking mass can be split into a non-negative term
containing part of the trace of the extrinsic curvature of $\Sigma$
in $M$ plus the remainder, that contains the trace of the extrinsic
curvature of $\mathcal{S}$ in $\Sigma$, say  $\mbox{tr}_\Sigma(k)$.
This remainder, which thus provides a lower bound for the Hawking mass,
is the Geroch energy, and it is defined explicitly as \cite{szabados_mass}
\[
  E_G(\mathcal{S})=\sqrt{\frac{Area(\mathcal{S})}{16\pi}}\left(1-\frac{1}{16\pi}\int_\mathcal{S}\left(\mbox{tr}_\Sigma(k)\right)^2d\mathcal{S}^2\right),
\]
where $d\mathcal{S}^2$ is the surface element of $\mathcal{S}$.
In the case of spheres on $\Sigma_{\text}$ the Geroch
and Hawking energies coincide
because $\Sigma_{\text}$ are minimal.
The application to spheres of constant $r$
on $\Sigma_{\tau}$ yields
\[
  E^\tau_G(r)=\frac{r_0}{2},
\]
which is a constant Geroch energy 
for all $r$ and equal to the ADM mass of $\Sigma_\tau$.
The remarkable point here is that the Geroch mass is a quasilocal quantity,
it does not depend on the asymptotic behavior (nor any other global property)
of the hypersurface ($\Sigma_\tau$ in this case).
Therefore, its constancy on all $\Sigma_\tau$ provides a
quasilocal characterization  of the parameter $r_0$.
This characterization of $r_0$ as some constant property on the
hypersurfaces $\Sigma_\tau$ plays an analogous role as $m$
being constant on the asymptotically flat hypersurfaces in Schwarzschild.

It is important to note that whatever notion of mass we choose
that mass is the same on all exterior regions of
the maximal extension $M$. This is in contrast
to what happens in many effective descriptions
of quantum spherical models in the literature,
where different exterior domains possess different masses
(see, e.g., \cite{Ashtekar:2018cay}).

It is also noticeable
that the level hypersurfaces of the function $\tau$ on $\mathcal{U}$,
which cross the transition hypersurface $\mathcal{T}$, have topology
$\mathbb{R}\times \mathbb{S}^2$. In fact, all spacelike slicings
in $M$ share that same topology, as opposed to what happens in
other geometrical constructions of nonsingular black holes,
in which different slicings can have
different topologies, see, e.g., \cite{Seno:2022_sings,Mars_1996,Maeda:2021jdc}.

All in all, in this section we have seen  that
the parameter $r_0$, being the minimum of the area of the orbits
of the spherical symmetry
can also be characterized in a global manner
by the difference
\begin{equation}\label{r0def}
    r_0=2 \lim_{r\to\infty}\big(M_{H}-M_K\big),
\end{equation}
as well as by the value of (twice) the ADM mass $M^\tau_{ADM}$ of the hypersurfaces $\Sigma_\tau$, and also quasilocally as (twice)
the value of the Geroch energy on \emph{any} sphere of constant $r$ on any $\Sigma_\tau$.
If the Einstein equations are satisfied,
then the limit on the right-hand side of \eqref{r0def} vanishes,
and therefore $r_0$ must also vanish as expected (see Sec.~\ref{sec:limits} below).

For completeness, the surface gravity $\kappa$, defined by  $\nabla_{\partial_\tau}(\partial_\tau)=\kappa\partial_\tau$ on each component $\horizon_\sigma$ of the horizon
with the Killing vector field $\partial_\tau$, reads
\[
  \kappa=\frac{\sigma}{4m}\sqrt{1-\frac{r_0}{2m}}.
\]
This satisfies the usual relation $|\kappa|=r^{-2}M_K|_{r=2m}$ (see, e.g., \cite{wald2010general}). Observe that in the limit case $r_0\to 2m$
we would obtain a vanishing surface gravity, in analogy with
the extremal Reisner-Nordstr\"om spacetime.
The appearance of a minimum area makes the surface gravity
smaller than that of a Schwarzschild black hole of mass $m$.

\subsection{A test scalar field propagating on the exterior region}\label{sec:qnm}
Although the transition surface $\mathcal{T}$ is well beyond the reach of an
  outside observer, the modifications performed to the theory through the polymerization \eqref{cantransf} and the linear combination of constraints \eqref{H_normal} have effects on the exterior regions.
  We have already seen how geometrically $r_0$ appears in quasilocal energy definitions.
  That information is carried out, of course, by the nonvanishing of the Ricci tensor
  everywhere, and that leaves traces that may be measurable from the asymptotic region
  by looking at effects on other physical fields. These
  effects, which could yield observational consequences, could
  be used to discard the model.
  As an example, we next consider the modification of the potential of a massless scalar field.

  The dynamics of such a test scalar field is given by the Klein-Gordon equation,
$$\Box\Phi=0.$$
By using $\text$ and $r_*$ \eqref{def:tortoise}
  as coordinates in the exterior region
  $D_S$ and
decomposing the scalar field in spherical harmonics,
\begin{align*}
\Phi(\text,r_*,\theta,\phi)=\frac{1}{r(r_*)}\sum_{l,m}\psi_l(\text,r_*) Y_{lm}(\theta,\phi),
\end{align*}
we can write the Klein-Gordon equation above as 
\begin{align*}
    \frac{\partial^2\psi_l}{\partial r_*^2}-\frac{\partial^2\psi_l}{\partial \tilde{t}^2} = V(r(r_*))\psi_l,
\end{align*}
with the potential term
\begin{align}\label{eq:qnm}
    V(\arguments) &= \left(1-\frac{2m}{r}\right)\frac{l(l+1)}{r^2}+\frac{1}{r}\frac{d^2r}{dr_*^2} 
 =\left(1-\frac{2m}{r}\right)\left(\frac{l(l+1)}{r^2}+\frac{4m+r_0}{2r^3}-\frac{3mr_0}{r^4}\right)\nonumber\\
    &=\left(1-\frac{2m}{r}\right)\left(\frac{l(l+1)}{r^2}-2\Psi_2-\frac{R}{6}\right).
\end{align}%
We can compare this potential with the one obtained for the Schwarzschild spacetime $V_{Sch}$
by means of their difference,
\begin{align*}
    V(\arguments)-V_{Sch}(r) &=V(\arguments)-\left(1-\frac{2m}{r}\right)\left(\frac{l(l+1)}{r^2}+\frac{2m}{r^3}\right)=\frac{r_0}{r^3}\bigg(1-\frac{2m}{r}\bigg)\left(\frac{1}{2}-\frac{3m}{r}\right), 
\end{align*}
that decays asymptotically as $r_0/r^3$. This difference is independent of $l$
and therefore all the modes are affected in the same way.

On the other hand, given that the appearance of $r_0$
in the geometry has consequences on the modifications of the mass
that could be interpreted as a Reissner-Nordstr\"om
geometry with mass $M_{RN}=m+r_0/2$ and charge $Q^2=2m r_0$,
we compare $V(r)$
with the corresponding potential $V_{RN}$.
Their difference is nonzero, and explicitly given by
\begin{align*}
       V(\arguments)-V_{RN}(r) &=V(\arguments)-\bigg(1-\frac{2m}{r}\bigg)\bigg(1-\frac{r_0}{r}\bigg)\left(\frac{l(l+1)}{r^2}+\frac{2m+r_0}{r^3}-\frac{4mr_0}{r^4}\right)\nonumber\\
     &=\frac{r_0}{r^3}\bigg(1-\frac{2m}{r}\bigg)\left( l(l+1)-\frac{1}{2}+\frac{3m+r_0}{r}-\frac{4mr_0}{r^2} \right),
\end{align*}
for which the leading term for any $l$ for large radius goes as $r_0/r^3$.
Note that, contrary to the previous case, the centrifugal term
in the potential $V(r)$ differs from that in Reissner-Nordstr\"om
with the aforementioned mass and charge.

This shows that the parameter $r_0$, and thus $\lambda$,
has a measurable imprint on the behavior of scalar fields
that differs from both Schwarzschild and Reissner-Nordstr\"om
far away from the horizon, in the asymptotic regions.
It is then to be expected that for more general fields
we may have a way to put this geometry to test.

\subsection{Schwarzschild and Minkowski limits}\label{sec:limits}
When $\lambda= 0$ we recover the Hamiltonian
formalism for vacuum in spherical symmetry in GR and, by definition,
$r_0$ vanishes. Conversely, we need $r_0=0$ to have an identically
vanishing Ricci tensor. Therefore  $r_0=0$
\emph{if and only if} the spacetime solves the equations of GR
for spherically symmetric vacuum,
and hence the \textit{limit} $r_0\to 0$ of
the solution with $m>0$
must clearly correspond to Schwarzschild with that $m$.
Checking this in
the static region $D_S$, where the metric reads \eqref{deformedschwarzsmetric},
is trivial, and we recover the usual form of the Schwarzschild metric
in one exterior domain.
In the homogeneous region  $D_h$, with the line element given by \eqref{metschwarzshom},
by setting $r_0=0$ we recover the usual interior geometry of the black-hole
(or white-hole) region in Schwarzschild. We must keep in mind that
now the range of the coordinate $T$ is given by $T\in(0,2m)$,
and that the curvature diverges as $T\to 0$ (see Sec. \ref{sec:curvature}). 
A nontrivial limit of spacetimes must be really taken if we consider
the covering domain $\mathcal{U}$ with the form of the metric \eqref{g:tau_x}.
When $r_0=0$ we have, from \eqref{rx}, that $r=|z|$.
The differentiability is lost at $z=0$, which makes \eqref{r-s-metric} ill defined there. In addition,
the positive lower bound of the function $r$ is lost
since it reaches $r=0$ and the transition
hypersurface $\mathcal{T}$ becomes unavailable (singularity). Therefore $\mathcal{U}$
must be split in two: one domain in which $z=r$
and another where $z=-r$. In the case $z=r$
we readily recover the Eddington-Finkelstein domain $D_{EF}$
with the metric \eqref{metricglp} with $r_0=0$ 
and the range $\rEF\in(0,\infty)$ (Gullstrand-Painlevé coordinates);
while in the case $z=-r$ we recover the equivalent domain
and metric up to a trivial change of sign in the crossed term
that amounts to an inversion of time.
Recalling \eqref{limit_gamma},
the metric \eqref{g:uv_ext} (for both values of $\sigma$)
on each respective extended domain $\mathcal{V}_\sigma$
corresponds to the usual null coordinates
of the Kruskal-Szekeres chart of Schwarzschild. 

Finally, the limit $m\to 0$ \emph{is Minkowski for any value of $\lambda$},
  and thus this limit holds in the full generalized Hamiltonian model.
  The limit follows from the fact that
  $m\to 0$ implies $r_0\to 0$ for any value of $\lambda$, 
and thus the limit $m\to 0$ of $(M,g)$ corresponds
to the limit $m\to 0$ of Schwarzschild, hence Minkowski, from above.

\section{Concluding remarks}\label{sec:conclusion}

It is known that the inclusion of holonomy corrections in the description of
nonhomogeneous models usually conflicts with the covariance of the theory.
In the present work, we have modified the Hamiltonian of spherical GR in such a way that it still obeys the Dirac hypersurface deformation algebra. We have done so through the canonical transformation \eqref{cantransf} and the linear combination with phase-space dependent coefficients \eqref{H_normal} of the GR constraints. These changes are regulated by a positive constant, the polymerization parameter $\lambda$. The limit $\lambda \to 0$
  leads to the GR Hamiltonian. In combination with the Dirac observable $m$, this polymerization
parameter defines a length scale $r_0$, which turns out to be fundamental in the geometrical description of the model.

We have then constructed the line element \eqref{metric_orig} associated to the modified Hamiltonian, so that the corresponding geometry is described covariantly, that is, coordinate changes coincide with gauge transformations on the phase space. We have performed a number of specific gauge choices that lead to different forms
of the metric tensor, see \eqref{deformedschwarzsmetric}, 
\eqref{metschwarzshom}, \eqref{g_I_full}, \eqref{g:tau_x}, and \eqref{metricglp},
and describe different regions of the spacetime. In particular, we have found a single chart $(\mathcal{U},g)$ that covers a globally hyperbolic interior homogeneous region and two exterior static regions, as depicted in Fig.~\ref{diagram:U}. We have produced the Penrose diagram of the domain $\mathcal{U}$
and have found the maximal analytic extension $M$ (see Fig.~\ref{diagram:maxextension}).

The classical singularity is fully resolved and replaced by a transition hypersurface $\mathcal{T}$, where the spheres attain their minimal area given
by 
$4\pi r_0^2$. This hypersurface is minimal and separates a trapped and an antitrapped region inside the horizon. Any free falling particle that crosses the black-hole horizon $\horizon_+$ from the exterior region $\exterior_+$ (see Fig.~\ref{diagram:maxextension}) enters the homogeneous interior region $\interior$, spends a finite amount of time there, and crosses the white-hole horizon $\horizon_-$ to emerge to the new asymptotically flat region $\exterior_-$.
All curvature scalars are finite everywhere and attain their maximum values on $\mathcal{T}$. Some eigenvalues of the Einstein tensor become negative, as one would expect for a singularity resolution. In fact, when defining the ``effective energy-momentum tensor'' $G^{a}{}_b$, all its corresponding standard energy conditions are violated everywhere except at the event horizon
  and at infinity, where they are all satisfied.

The effects produced
  by the polymerization of the Hamiltonian and the linear combination of the constraints resolve the singularity inside the black-hole horizon, but also leave specific traces in the exterior region. In order to see this in more detail, we have computed 
the energy content of the spacetime through different definitions of mass, and we have also considered a test scalar field propagating on this background spacetime.
Let us remark that the present work studies a symmetry-reduced vacuum model and, therefore, a phenomenological analysis (say on the lines of, e.g., \cite{Barrau:2018rts}) of this solution would require full control over a perturbation scheme outside spherical symmetry, which is beyond the scope of this work.

Concerning the masses, contrary to what happens in GR, the Hawking and Komar masses do not coincide since the Einstein equations do not hold.
Both depend on the area-radius function, but change in a different way due to the presence of quantum-gravity effects parametrized by $r_0$. The Hawking mass coincides with the constant of motion $m$ on the horizon, whereas the Komar mass attains this precise value at infinity. The difference of these two masses at infinity turns out to be $r_0/2$.
Furthermore, the fall-off conditions of the Ricci scalar turns out to be enough to define the ADM mass \eqref{eq.adm}, which is a nontrivial feature for modified theories of gravity (see, e.g., \cite{New_Masses}).
We note also that the limit $m\to 0$,
for any value of $\lambda$, leads to Minkowski in the full generalized Hamiltonian model.

Regarding the test scalar field, it has been assumed to obey the Klein-Gordon equation. After decomposing it in spherical harmonics, its potential \eqref{eq:qnm} has been obtained and compared with the corresponding potential for Schwarzschild and a compatible Reissner-N\"ordstrom backgrounds.

To finish the conclusions,
let us comment about the relation between the polymerization constructed here
and others that have been presented in the literature. In particular, in the context of homogeneous models, there are three main strategies to deal with the polymerization parameter. It might be considered to be constant over the whole phase space (the so-called $\mu_0$-scheme), to be constant on solutions, or to depend on some of the triad components ($\bar{\mu}$-scheme). Each of these approaches has advantages and drawbacks, mainly regarding the confinement of quantum effects to large-curvature regions. As commented above, in
the present model we have the constant polymerization parameter $\lambda$ and the fundamental
length scale $r_0$, which are related by relation \eqref{r0def} in terms of the constant of motion
$m$. Therefore, the first two schemes are included in the present model, but not the $\bar{\mu}$-scheme. More precisely, in the $\mu_0$-scheme
$\lambda$ would be a universal constant (on the whole phase space), and thus $r_0$ would change linearly with the mass $m$ of the black hole.
In contrast, if we define $r_0$ to be a universal length scale, independent of the specific solution, then the polymerization parameter $\lambda$ would be
a constant on each solution, but would change with the mass $m$. In any case, all the presented features of the model turn out to be insensitive
to the particular choice between these two schemes and, in particular, the transition surface is always hidden inside the horizon $(r_0<2m)$.

\begin{acknowledgments}
We thank Martin Bojowald, Marc Mars, Javier Olmedo, and Jos\'e M. M. Senovilla for fruitful discussions and interesting
comments. We acknowledge financial support from the Basque Government Grant \mbox{No.~IT956-16}
and from the Grant FIS2017-85076-P, funded by
MCIN/AEI/10.13039/501100011033 and by “ERDF A way of making Europe”.
AAB is funded by the FPI fellowship \mbox{PRE2018-086516} of the Spanish MCIN.
\end{acknowledgments}

\appendix

\section{Alternative line elements and the Bojowald-Brahma-Yeom model}\label{app.bby}

Let us consider the following variation of the line element \eqref{metric},
\begin{equation}\label{rescaledmetric}
 ds^2=-f(t,x) N(t,x)^2 dt^2+\frac{f(t,x)}{F(t,x)}(dx+N^x(t,x) dt)^2+\erad(t,x) d\Omega^2,
\end{equation}
$f(t,x)\neq 1$ being an arbitrary scalar function.
This line element does not obey the condition
{\it i/} from Sec. \ref{sec.lineelements}
but, as we are going to show below, it is covariant in the sense
of condition {\it ii/}, i.e., gauge transformations correspond to coordinate transformations.

Since $f$ is a scalar, it transforms as
\begin{equation}
 \delta f= \dot{f} \xi^t+f' \xi^x.
\end{equation}
Therefore, under a coordinate transformation of \eqref{rescaledmetric}
the change of $\overline N:=\sqrt{f}N$ is in fact the same
as \eqref{coordN}, that is,
\begin{eqnarray}
\delta \overline{N} &=& \dot{\overline{N}} \xi^t+ \overline{N}' \xi^x
+\overline{N} (\dot{\xi}^t-N^x \xi^{t\prime}).
               \label{coordN2}
 \end{eqnarray}
The new spatial component of the metric $\overline{q}_{xx}:=f/F$ changes
in the same way as \eqref{coordqxx},
\begin{eqnarray}
 \delta \overline{q}_{xx} &=& \dot{\overline{q}}_{xx} \xi^t+\overline{q}_{xx}' \xi^x+2\overline{q}_{xx} (N^x  \xi^{t\prime}+\xi^{x\prime}),
 \end{eqnarray}
whereas the change of $N^x$ is given by
\begin{eqnarray}
  \delta N^{x} &=& \dot{N}^x \xi^t+N^{x\prime} \xi^x+N^x
    (\dot{\xi}^t-\xi^{x\prime})-\left[\frac{\overline{N}^2}{\overline{q}_{xx}} +(N^x)^2\right] \xi^{t\prime}+\dot{\xi}^x \nonumber\\\label{coordNx2}
&=& \dot{N}^x \xi^t+N^{x\prime} \xi^x+N^x
    (\dot{\xi}^t-\xi^{x\prime})-\left[FN^2 +(N^x)^2\right] \xi^{t\prime}+\dot{\xi}^x.
    \end{eqnarray}
If one then assumes that the Hamiltonian of the system is $H[N]+D[N^x]$,
the gauge transformations of the lapse $N$ and the shift $N^x$ are given by
\eqref{gaugeN} and \eqref{gaugeNx}, respectively,
which fit with \eqref{coordN2}
and \eqref{coordNx2}.

In summary, instead of \eqref{metric}, one may consider the metric \eqref{rescaledmetric} to describe the geometry of the model in the sense that gauge transformations
  in phase space correspond to spacetime coordinate transformations. However,
the price to pay to use \eqref{rescaledmetric} is that the time vector $\partial_t$
(used to perform the canonical analysis) is modified and, in particular,
the normal vector $n$ ceases to be unit for
the deformed metric.

In Ref. \cite{Bojowald:2018xxu} 
a line element of the form \eqref{rescaledmetric} is presented
to provide the geometric interpretation of a different polymerized canonical system.
However, the problem is that the corresponding factor $f$ finally chosen
to explicitly work out the model is not a scalar quantity.
More precisely,
in the polymerized model constructed in \cite{Bojowald:2018xxu} the structure
function that appears in the bracket between two Hamiltonian constraints is given
by $F=\beta(\kang){E}^x/({E}^\varphi)^2$, 
whereas the considered line element is
\begin{align}\label{bbylineelement}
    {ds}^2_{BBY} = -\beta(\kang)\lapse^2 {dt}^2 + \frac{({E}^\varphi)^2}{{E}^x}({dx}+\shift{dt})^2 +{E}^x{d\Omega}^2,
\end{align}
which corresponds to 
\eqref{rescaledmetric}
by setting $f=\beta$. But, since $\kang$ is not a scalar
  quantity, 
$\beta(\kang)$
does not transform as a scalar either, and the line element
\eqref{bbylineelement} turns out not to be covariant.
In the following we show this fact explicitly
by presenting the solution to the equations of the model of Ref. \cite{Bojowald:2018xxu}
in two different gauges.
We will construct the line element \eqref{bbylineelement} for each gauge choice
and show that they can not be related by a coordinate transformation
since they provide two different families of geometries: one with a vanishing
and the other with a nonvanishing Ricci scalar.

The evolution equations to be considered
are eqs. (19)--(22) in Ref. \cite{Bojowald:2018xxu}, with the specific choice 
(15)--(16) for the free functions, which in turn makes $\beta=\cos(2\delta\kang)$
with $\delta$ a nonvanishing
polymerization parameter.
The first gauge choice is the one already considered in \cite{Bojowald:2018xxu} for the exterior
region, that is ${E}^x=x^2$ and ${K}_\varphi=0$, which
provides the Schwarzschild metric [see eq.~(63) in \cite{Bojowald:2018xxu}],
\begin{align}\label{bby1}
    {ds}^2_{BBY(1)} = -\left(1-\frac{2M}{x}\right){dt}^2 + \left(1-\frac{2 M}{x}\right)^{-1}{dx}^2 +x^2{d\Omega}^2,
\end{align}
for a constant $M$. This metric, of course, has a vanishing Ricci scalar $R_{BBY(1)}=0$.

The second gauge choice
consists on taking ${E}^x=x^2$ and ${E}^\varphi=x$.
The general solution of eqs. (19)--(22) in \cite{Bojowald:2018xxu}
is then given by
\[
  K_x=\frac{\epsilon}{2x}\sqrt{\frac{2\widetilde{M}}{x}}\left({1-\frac{2\widetilde{M}\delta^2}{x}}\right)^{-1/2}, \quad
  \sin^2(\delta K_\varphi)=\frac{2\widetilde{M}\delta^2}{x},\quad
  N=N_0,\quad N^x={\epsilon N_0}\sqrt{\frac{2 \widetilde{M} }{x}}
  \sqrt{1-\frac{2\widetilde{M}\delta^2}{x}},
\]
where we have defined $\epsilon:=-\sgn(\sin(2\delta K_\varphi))$, 
$N_0$ and $\widetilde M$ are integration constants, whereas
$x$ is restricted to have the same sign as $\widetilde M$ and to obey
$2\delta^2|\widetilde{M}|<|x|$.
Therefore, for this gauge, the line element
\eqref{bbylineelement}
takes the form,
\begin{align}\label{bby2}
    {ds}^2_{BBY(2)} = -\left(1-\frac{2\widetilde M}{x}-\frac{4\widetilde M\delta^2}{x}\left(1-\frac{\widetilde M}{x}\right)\right)
    d\tau^2 +2\sqrt{\frac{2\widetilde M}{x}\left(1-\frac{2\widetilde M\delta^2}{x}\right)}d\tau dx + dx^2 +x^2d\Omega^2,
\end{align}
where, without loss of generality, $N_0$ and $\epsilon$ have been absorbed in the definition of $\tau$.
This line element presents Euclidean $(2\delta^2 |\widetilde M|<|x|<4\delta^2 |\widetilde M|)$
 and Lorentzian $(4\delta^2 |\widetilde M|<|x|)$ regions, with Ricci scalar
\begin{align}
    R_{BBY(2)} = 4\widetilde{M}^2\delta^2\frac{2\delta^2(x+2\widetilde{M})-3x}{x^3(x-4\widetilde M \delta^2)^2},
\end{align}
which only vanishes identically in the case $\widetilde M=0$ that corresponds to Minkowski.
Therefore, we conclude that for $\delta\neq 0$, even in the Lorentzian regions, 
there is no coordinate transformation that relates the line elements \eqref{bby1} and \eqref{bby2},
and thus they correspond to two different metric tensors.
As a result, two different gauge choices in the construction
  provide two different geometries and the line element
  \eqref{bbylineelement} is not covariant, that is, it does
  not satisfy condition \textit{ii/} above.

\bibliography{refs}

\end{document}